\documentclass[oneside, sn-mathphys-num]{sn-jnl} 
\usepackage{graphicx}%
\usepackage{multirow}%
\usepackage{amsmath,amssymb,amsfonts}%
\usepackage{amsthm}%
\usepackage{mathrsfs}%
\usepackage[title]{appendix}%
\usepackage{xcolor}%
\usepackage{textcomp}%
\usepackage{manyfoot}%
\usepackage{booktabs}%
\usepackage{algorithm}%
\usepackage{algorithmicx}%
\usepackage{algpseudocode}%
\usepackage{listings}%
\usepackage{rotating}
\usepackage{pdflscape}
\usepackage{booktabs}
\usepackage{comment}

\newcommand{\lat}{\textit{Fermi}/LAT}
\newcommand{\gbm}{\textit{Fermi}/GBM}

\theoremstyle{thmstyleone}%

\theoremstyle{thmstyletwo}%

\theoremstyle{thmstylethree}%

\begin{document}
\title[GRB 260226A]{The ultra-fast afterglow of GRB\,260226A}

\author*[1,2]{\fnm{Biswajit} \sur{Banerjee}}\email{biswajit.banerjee@gssi.it}
\author*[3]{\fnm{Alessio} \sur{Mei}}\email{alessio.mei@inaf.it}
\author*[1,2]{\fnm{Annarita} \sur{Ierardi}}\email{annarita.ierardi@gssi.it}
\author[4]{\fnm{Samanta} \sur{Macera}}
\author[1,2]{\fnm{Gor} \sur{Oganesyan}}
\author[5]{\fnm{Shraddha} \sur{Mohnani}}
\author[6]{\fnm{Elias} \sur{Kammoun}}
\author[1,2]{\fnm{Pawan} \sur{Tiwari}}
\author[1,2]{\fnm{Stefano} \sur{Ascenzi}}
\author[1,2]{\fnm{Samuele} \sur{Ronchini}}
\author[1,2]{\fnm{Ansh} \sur{Chopra}}
\author[1]{\fnm{Alessio Ludovico} \sur{De Santis}}
\author[3,7]{\fnm{Stefano} \sur{Covino}}
\author[3]{\fnm{Paolo} \sur{D'Avanzo}}
\author[4]{\fnm{Andrea} \sur{Melandri}}
\author[4]{\fnm{Silvia} \sur{Piranomonte}}

\affil[1]{Gran Sasso Science Institute (GSSI), I-67100 L'Aquila, Italy}
\affil[2]{INFN, Laboratori Nazionali del Gran Sasso, I-67100 Assergi, Italy}
\affil[3]{INAF – Osservatorio Astronomico di Brera, Via E. Bianchi 46, 23807 Merate (LC), Italy}
\affil[4]{INAF - Osservatorio Astronomico di Roma, Via Frascati 33, I-00078, Monte Porzio Catone, Italy}
\affil[5]{Department of Astronomy, Astrophysics and Space Engineering, Indian Institute of Technology Indore, India}
\affil[6]{Cahill Center for Astronomy \& Astrophysics, California Institute of Technology, 1216 East California Boulevard, Pasadena, CA 91125, USA}
\affil[7]{Como Lake centre for AstroPhysics (CLAP), DiSAT, Università dell’Insubria, via Valleggio 11, 22100 Como, Italy}

\abstract{Long-duration gamma-ray bursts are typically powered by relativistic jets launched after the core collapse of some rapidly rotating massive stars. Internal dissipation releases part of the jet energy as highly variable MeV prompt emission, while the remaining kinetic energy drives an external shock into the surrounding medium and produces the so-called afterglow. 
During the first minutes of the afterglow, the unsteady jet transfers energy to the external shock. The early afterglow emission in MeV-GeV energies is rarely observed because the emergence of afterglow can be overshined by the prompt emission.
Here we report exceptional observations of GRB\,260226A with the \textit{Fermi} Large Area Telescope, which recorded the largest number of photons above 100 MeV from a gamma-ray burst. These data allow us to reconstruct the evolution of the bolometric flux of the afterglow from its emergence during the prompt emission phase with unprecedented detail.
The afterglow component peaks near \(50\) MeV and fades rapidly, first as \(t^{-1.5}\) and then transiting to an ultra-fast \(t^{-2.8}\) decay after about one minute.
This behavior cannot be explained by standard synchrotron emission from a blast wave propagating into a cold medium. We interpret it as external inverse Compton radiation from freshly heated electrons cooling on prompt photons in a dense, pair-loaded stellar wind. GRB\,260226A therefore shows that MeV--GeV observations can directly reveal the formation of the external shock and the massive-star environment is significantly reshaped by the prompt emission.}

\maketitle
Gamma-ray bursts (GRBs) are extragalactic keV--MeV transients that release isotropic-equivalent energies of \(E_{\rm iso}\sim10^{50}-10^{55}\,{\rm erg}\) on timescales of seconds to minutes. They are powered by ultra-relativistic, collimated jets launched either after the core collapse of some fast-rotating Wolf-Rayet stars \citep{Woosley1993} or compact binary mergers \citep{Blinnikov1984,Eichler1989}. The observed radiation is divided into two phases. The prompt emission consists of brief energetic MeV flashes with a duration of \(\sim0.1\) - \(100\) s, produced by dissipation within the relativistic jet \citep{Rees1994,Sari1997}. The afterglow is a longer-lived broadband emission, observed from radio wavelengths to TeV gamma-rays \cite{MAGIC:2019lau, HESS:2021dbz, Abdalla:2019dlr, LHAASO:2023kyg,Abe:2023nhj}, generated as the jet transfers its kinetic energy to the surrounding medium and drives a relativistic blast wave \citep{Paczynski1993,Meszaros1997,Sari1998}. The prompt emission phase therefore probes dissipation inside the jet, whereas the afterglow probes the circumburst environment and the microphysics of relativistic shocks.

The transition between the prompt and afterglow phases remains poorly explored. During this phase, the unsteady jet begins to establish an external shock, and therefore contains direct information on how jet energy is transferred to the circumburst medium \citep{Beloborodov2014,DP2024}. However, observationally, this transition is difficult to isolate. At keV--MeV energies, where the prompt emission is brightest, the emerging afterglow is usually 
overshined by the fading prompt emission. At higher energies, the prompt emission contribution is weaker, and radiation from the most energetic particles in the external shock can become visible \citep{Ghisellini2010,Ghirlanda2010,Kumar2010,Nava2014}. 
\begin{figure}
\centering
\includegraphics[width=\textwidth]{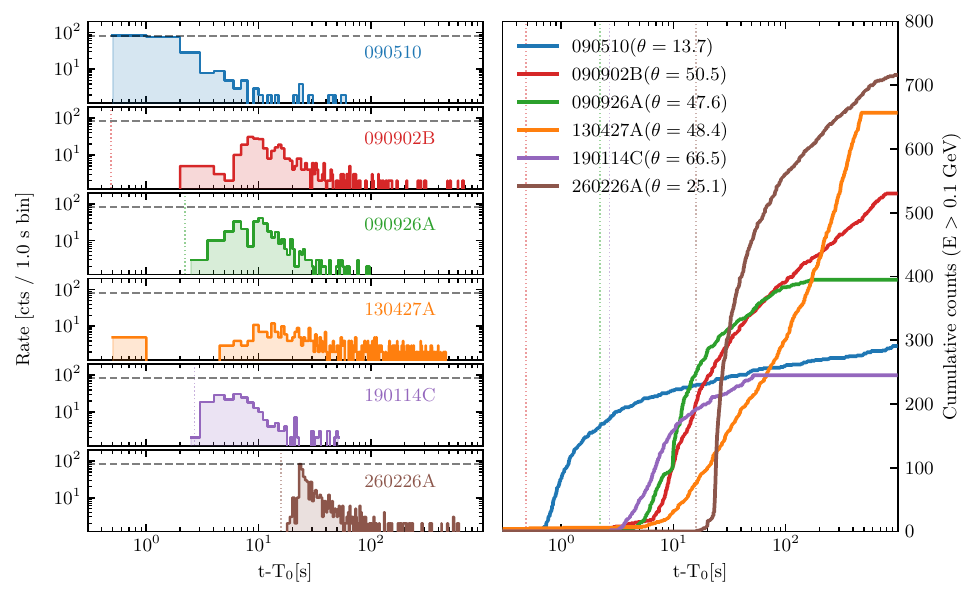}
\includegraphics[width=\textwidth]{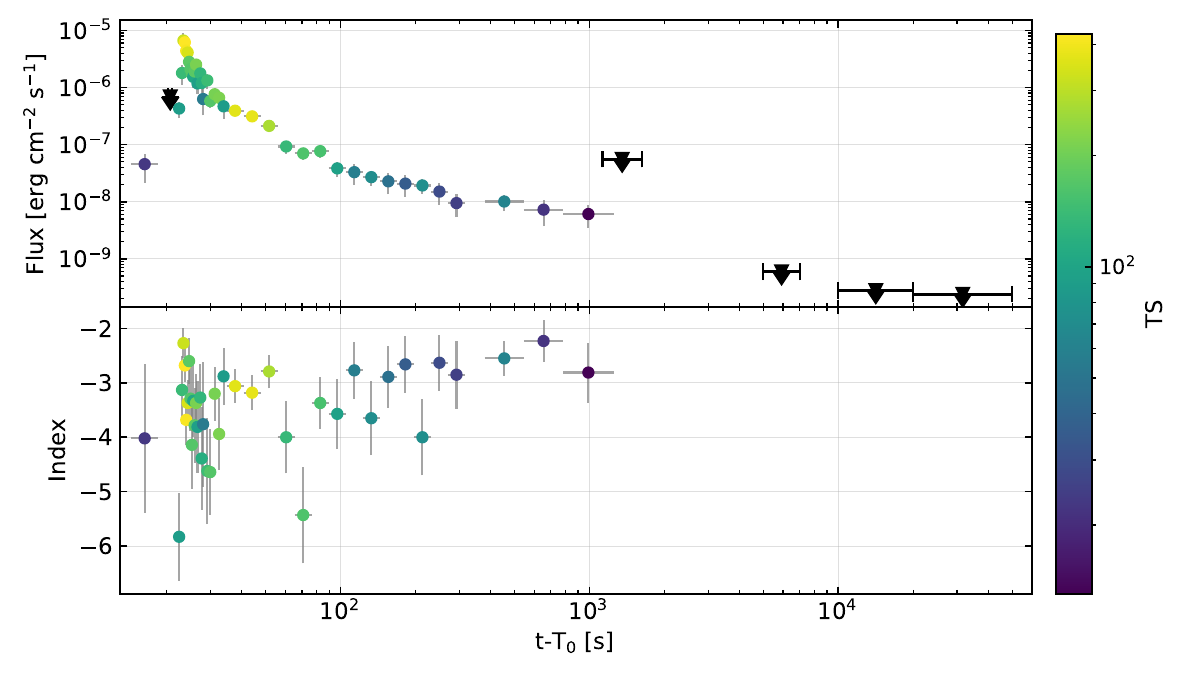}
\caption{
GeV emission of GRB\,260226A compared with the previous brightest \textit{Fermi}/LAT bursts from the second \textit{Fermi}/LAT GRB catalog \citep{Ajello:2019zki}.
\textit{Upper left:} LAT count-rate light curves with 1 s binning for GRB\,260226A and GRBs 090510, 090902B, 090926A, 130427A and 190114C. Vertical dotted lines mark the LAT onset relative to the GBM trigger, and horizontal dashed lines show the peak count rate of GRB\,260226A.
\textit{Upper right:} Cumulative LAT counts above 0.1 GeV. GRB\,260226A reaches more than 700 photons within the first \(10^3\) s. The parameter $\theta$ indicates the boresight angle, namely, the angle between the source position and the LAT boresight.
\textit{Lower panels:} Time-resolved LAT \(0.1\)--\(1\) GeV flux and photon index for GRB\,260226A. Flux points are colored by the test statistic from the likelihood analysis; triangles mark 95\% confidence upper limits for bins with TS \(<9\). The high-energy emission peaks around \(T_0+23\)--\(25\) s and then fades rapidly.
}
\label{fig:lat_lc}
\end{figure}
The Large Area Telescope (LAT; \(0.1\)--\(10\) GeV; \cite{2009ApJ...697.1071A}) onboard the \textit{Fermi} Gamma-Ray Telescope has detected high-energy emission above \(100\) MeV from several GRBs simultaneously with their keV--MeV prompt emission. However, photon statistics at GeV energies are often limited, and the physical origin of early GeV emission in GRBs remains debated \citep{Ghisellini2010,Ghirlanda2010,090510,Kumar2010,Beloborodov2014,Vurm2014,Hascoet2015,190114C,Ravasio2019,Macera:2025wrv,Maraventano2026}. These observations show that GeV radiation often begins delayed with respect to the MeV prompt emission phase and is frequently described as an additional hard spectral component superimposed on the prompt emission spectrum \citep{Ghirlanda2010,090510,090902B,090926A,190114C,Ravasio2019,Macera:2025wrv}. The interpretation of this delayed GeV component is still uncertain: it may represent high-energy prompt emission, the onset of the external-shock afterglow, or a combination of both. The distinction requires broadband spectral coverage across the MeV gap, especially between \(\sim10\) and \(100\) MeV, where the prompt and afterglow components can be separated most clearly. The Energetic Gamma Ray Experiment Telescope (EGRET) has detected bright counterpart of GRB 941017 in the MeV gap of 10-200 MeV \cite{Gonzalez2003}. Dedicated LAT Low Energy (LLE; $30-100$ MeV) analyses have recovered variable high energy emission in a small number of GRBs \citep{Pelassa2010,Vianello2011,Chand2020,Mei2022,Ravasio2024, Ajello:2019zki}, but afterglow emission in this band remains difficult to measure because of the limited sensitivity and large point spread function of the photons. As a result, constructing a high-energy broadband afterglow spectrum from \(\sim10\) MeV to \(\sim100\) MeV energies has not been possible.

GRB\,260226A was detected on February 26 2026 at 10:37:55 UTC (considered as trigger time T$_{0}$) by the Gamma-Ray Burst Monitor (GBM; $8\ {\rm keV} -40\ {\rm MeV}$) on board \textit{Fermi} \citep{Meegan2009,2026GCN.43840....1F,2026GCN.43851....1B}. The exceptionally bright GeV emission also triggered the \textit{Fermi}/LAT onboard seeded-alert system, making it only the second GRB in the history of the mission, after GRB\,090510, to produce an autonomous LAT alert \citep{Ackermann2010,2026GCN.43844....1D}. Compared with the brightest LAT GRBs reported in the second LAT GRB catalog \cite{Ajello:2019zki}, GRB\,260226A yielded more than 700 photons above \(100\) MeV within the first \(10^3\) s, the largest number recorded from a GRB to date (Fig.~\ref{fig:lat_lc}). The light curve in the energy range of \(0.1\)--\(1\) GeV shows a bright early flare followed by a shallower decline. The redshift of GRB~260226A is not known, due to the absence of a firm optical afterglow detection despite prompt and deep follow-up (e.g., $r > 20.4$ AB at 7.77 hr \cite{reguitti2026grb} and deeper non-detections from COLIBR{\'I} and LCOGT \cite{angulo2026grb,strausbaugh2026grb}).

In this work, we present a time-resolved broadband spectral analysis of GRB\,260226A using publicly available \textit{Fermi} data. We combine \textit{Fermi}/GBM observations from \(\sim10\) keV to \(40\) MeV with standard LAT data above \(100\) MeV. To close the spectral gap between these instruments, we develop a dedicated analysis of low-energy LAT events in the \(30\)--\(100\) MeV range (see Methods). This allows us to reconstruct the continuous keV--GeV spectral evolution of GRB\,260226A over about the first 10 minutes after trigger, covering the prompt emission, the emergence of the high-energy afterglow, and its subsequent rapid decline.
\begin{figure}
\centering
\includegraphics[width=1\textwidth]{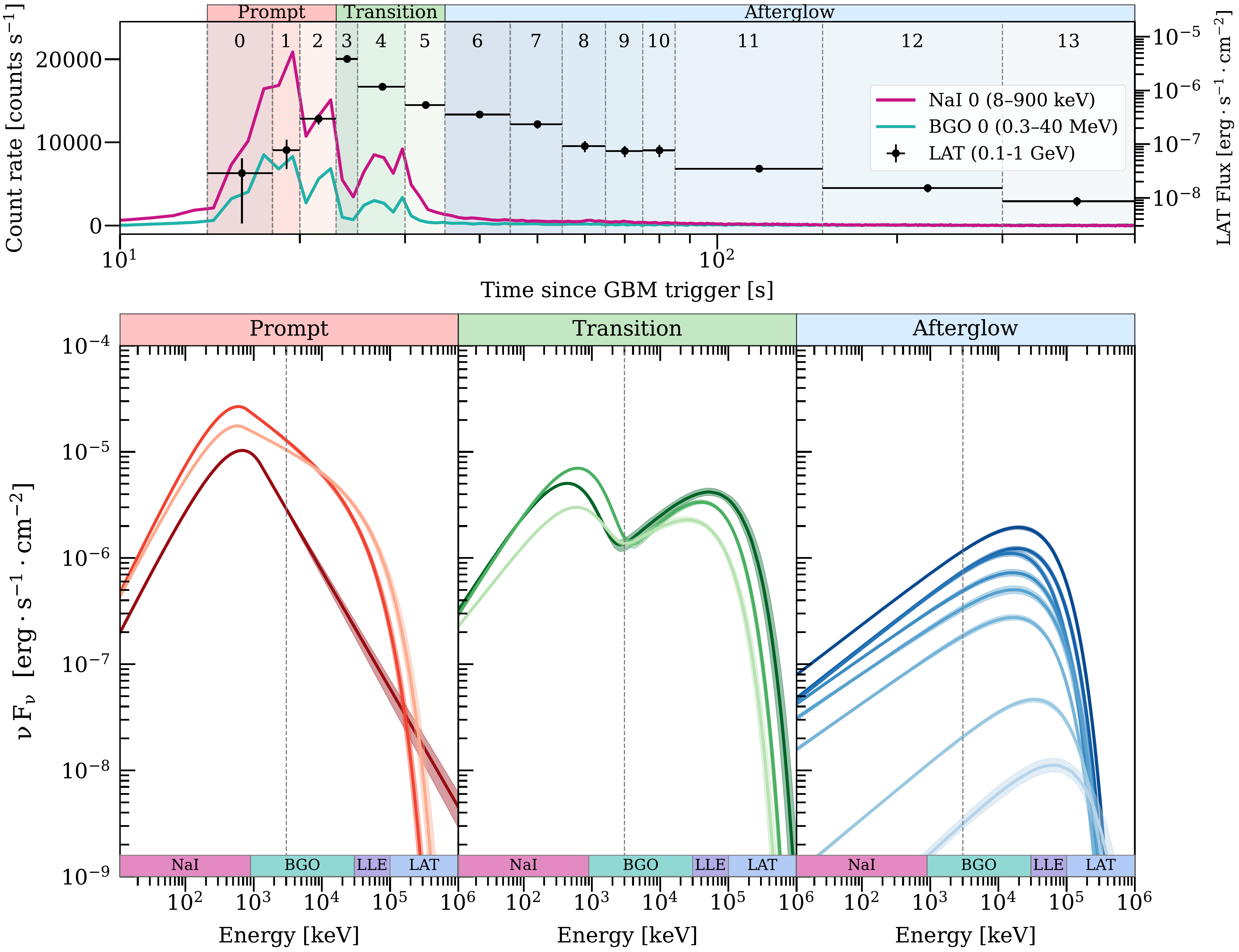}
\caption{
Broadband time-resolved spectral evolution of GRB\,260226A.
\textit{Upper panel:} \gbm\ count-rate light curve and \lat\ flux light curve. Vertical dashed lines delimit the spectral time bins. Prompt, transition and afterglow-dominated intervals are shown in red, green and blue, respectively.
\textit{Lower panel:} Best-fit models and \(1\sigma\) confidence regions for the prompt, transition and afterglow-dominated spectra. The dashed vertical line at 3 MeV indicates the approximate energy separating the domains where the prompt and high-energy afterglow components dominate. Coloured bands indicate the main energy ranges covered by NaI, BGO, LLE and LAT data used in the fits.
}
\label{fig:time_res_spectral_analysis+LC}
\end{figure}

\begin{figure}
\centering
\includegraphics[width=1\linewidth]{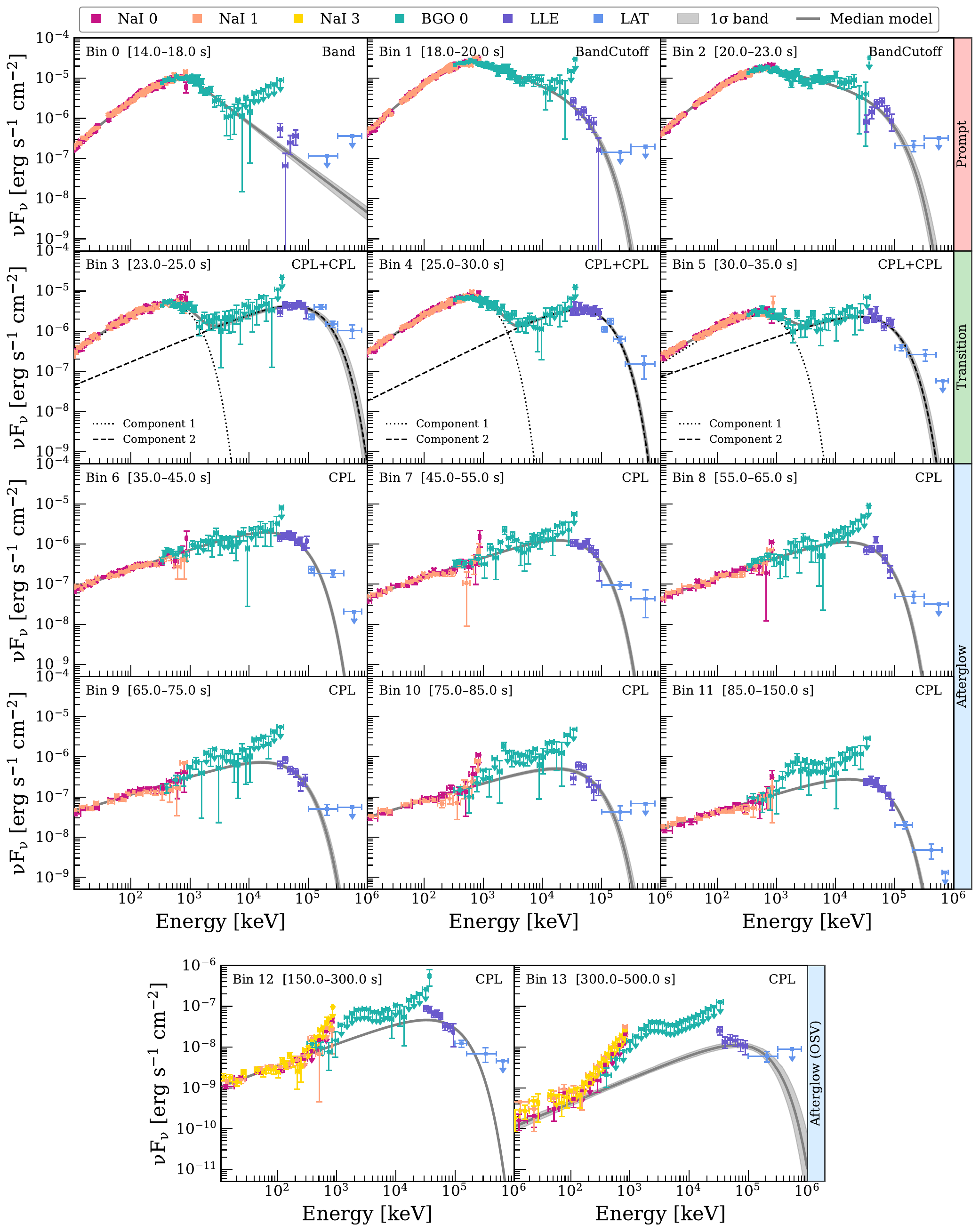}
\caption{
Individual time-resolved spectra of GRB\,260226A. The first row shows the prompt emission phase, the second row the transition phase, and the following rows the afterglow-dominated phase, including the two late spectra analysed with the orbital-subtraction background method. Data points and \(1\sigma\) uncertainties are shown for the NaI, BGO, LLE and LAT datasets; arrows indicate \(3\sigma\) upper limits. The best-fit model is shown by the solid black line, with its \(1\sigma\) confidence region in gray. In the transition-phase spectra, dotted and dashed lines show the prompt and high-energy afterglow components separately.
}
\label{fig:time_res_individual_spectra}
\end{figure}
\section*{Results}

\subsection*{Broadband spectral evolution}

We divided the burst into 14 temporal bins between \(T_0+14\) s and \(T_0+500\) s, depending on the morphology of the MeV count-rate light curve and the GeV flux light curve (Fig.~\ref{fig:time_res_spectral_analysis+LC}). For the final two bins, covering \(T_0+150\)--\(500\) s, we constructed a dedicated background model using preceding and subsequent \textit{Fermi} orbits, as described in the Methods. Spectral models were compared using the Akaike Information Criterion \citep{Akaike1974}, and parameter uncertainties were derived from Markov chain Monte Carlo sampling.

The combined temporal and spectral behavior separates the burst into three phases (Figs.~\ref{fig:time_res_spectral_analysis+LC} and \ref{fig:time_res_individual_spectra}). The prompt emission light curve is dominated by a bright double-pulsed MeV structure up to \(T_0+23\)~s (Fig.~\ref{fig:time_res_spectral_analysis+LC}, top panel), while the GeV emission begins to emerge. From \(T_0+23\) s to \(T_0+35\)~s, the prompt MeV component fades rapidly as the high-energy emission reaches its maximum, marking the overlap between prompt emission and the emerging afterglow in the so-called transition phase. After \(T_0+35\) s, the spectra are dominated by a single high-energy afterglow component that persists to at least \(T_0+1.0\) ks. 

The spectral evolution reveals a clear distinction between the prompt and afterglow components. 
The prompt emission spectra are well described by a Band function \cite{Band1993} peaking at $E_{\rm p}\sim0.5$ MeV (top row of Fig.~\ref{fig:time_res_individual_spectra}), which evolves into a cutoff power-law during the transition phase (second row of Fig.~\ref{fig:time_res_individual_spectra}). In contrast, the afterglow component is consistently well described by a cutoff power-law, with a spectral peak at $\sim15$--$60$ MeV, from its emergence during the transition phase and throughout the subsequent observations (second to last rows of Fig.~\ref{fig:time_res_individual_spectra}).
When the afterglow-dominated spectra are instead fitted with a Band function, the high-energy photon index resulted in very soft values \(\beta<-3\). This clearly indicates the absence of a significant high-energy power-law tail in the spectrum and motivates the use of a cutoff power-law. The two components also differ below their peaks: prompt emission has the usual hard photon index, \(\alpha\simeq-1\), whereas afterglow remains systematically softer, with \(\alpha\simeq-1.5\) throughout its observed evolution (see Fig.~\ref{fig:time_res_individual_spectra}).

\subsection*{Bolometric evolution of the afterglow}

\begin{figure}
\centering
\includegraphics[width=0.9\textwidth]{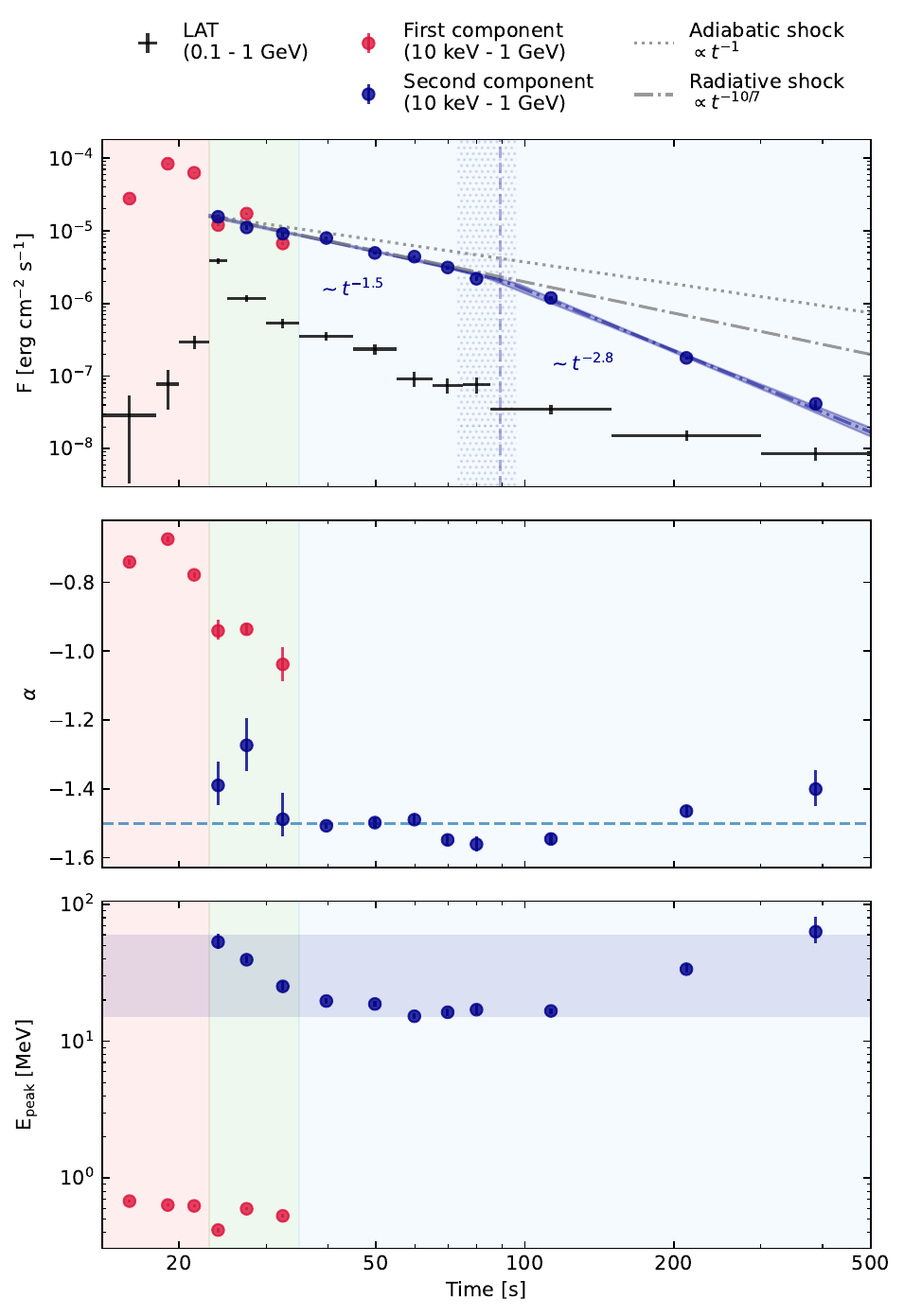}
\caption{
Temporal evolution of the first component (prompt emission, red dots) and second component (afterglow, blue dots). Background colours mark the prompt, transition and afterglow-dominated phases defined in Fig.~\ref{fig:time_res_spectral_analysis+LC}.
\textit{Upper panel:} Bolometric \(10\) keV--\(1\) GeV light curve. The second component is fitted with a broken power law. The dashed curve shows the posterior median and the shaded region shows the \(1\sigma\) interval. The break time and its uncertainty are marked by the vertical dashed line and shaded band. Black symbols show the LAT-only \(0.1\)--\(1\) GeV flux for comparison.
\textit{Middle panel:} Low-energy photon index of the spectral model. The dashed line marks \(\alpha=-1.5\).
\textit{Lower panel:} Evolution of the spectral peak energy. The horizontal band marks the 15--60 MeV range, where the peak energy remains throughout the emission.
}
\label{fig:second_component}
\end{figure}

We characterized the spectral and temporal evolution of the high-energy afterglow component. Throughout the observations, the low-energy photon index remains close to \(\alpha\simeq-1.5\) (Fig.~\ref{fig:second_component}, middle panel), indicating a stable spectral shape despite the rapid fading of the flux. The peak energy evolves non-monotonically between 15-60 MeV (Fig.~\ref{fig:second_component}, lower panel). 

The bolometric afterglow light curve (\(10\,{\rm keV}\)--\(1\,{\rm GeV}\)) is well described by a broken power law (Fig.~\ref{fig:second_component}, upper panel). During the first minute, the flux decays as \(t^{-1.5}\). After a break at \(T_0+65\) s, the decline steepens to an ultra-fast \(t^{-2.8}\) decay.

The bolometric afterglow flux is one to two orders of magnitude larger than the \(0.1\)--\(1\) GeV flux inferred from standard LAT analysis alone, shown by the gray symbols in Fig.~\ref{fig:second_component}. Thus, in GRB\,260226A the standard LAT band does not trace the bulk of the high-energy afterglow power: most of the energy is emitted below \(100\) MeV, in the usually poorly sampled MeV gap between 10 and 100 MeV. In addition, the nominal MeV duration of the burst, \(\sim80\) s, does not entirely correspond to prompt emission. Our spectral decomposition shows that the classical prompt component dominates only during the first \(\sim35\) s.

This bolometric reconstruction changes the inferred temporal behavior of the high-energy afterglow. Several LAT-detected GRBs show GeV light curves that evolve from a steep early decay to a shallower late-time decline \citep{Ghisellini2010,Ghirlanda2010,Ackermann2010}. However, in GRB \,260226A, the bolometric keV--GeV light curve shows the opposite behavior: a \(t^{-1.5}\) decay followed by a much steeper \(t^{-2.8}\) decline. This difference is revealed only when the MeV data are included, demonstrating that GeV-only light curves may not include the dominant energy output of the early high-energy afterglow.

\section*{Discussion}
The rapid temporal decay and unusual spectral shape of the high-energy afterglow challenge the standard external shock interpretation. In the standard afterglow model, the kinetic energy remaining after prompt emission is transferred from the relativistic ejecta to the forward shock. Once this energy transfer is complete, the forward shock approaches the self-similar Blandford--McKee deceleration regime \citep{BM1976}. In the adiabatic limit, which successfully describes much of the late-time multi-wavelength afterglow population, the bolometric luminosity is expected to decline as \(t^{-1}\) \citep{Granot&Sari2002}. Even in the extreme fully radiative limit, the expected decline is only moderately steeper, close to \(t^{-10/7}\) \citep{Katz1997,Vietri1997,Ghisellini2010}. 
Instead, the high-energy bolometric afterglow of GRB \,260226A steepens as \(\sim t^{-2.8}\) after about one minute, too rapidly to be explained by a standard blast wave propagating in a cold medium.

In addition to the steepening of the light curve, the spectrum provides an independent proof of incompatibility with standard afterglow powered by synchrotron radiation of shock-accelerated electrons, further constraining the radiation mechanism (see Methods for more details). A synchrotron origin for a component peaking at \(E_{ p}\simeq40\)--\(50\) MeV would require the characteristic forward shock electrons to radiate in comoving magnetic fields of order \(B'\sim10^{5}\) G, many orders of magnitude larger than expected from shock-amplified fields in either a uniform medium or a stellar wind. Associating the peak with the highest-energy synchrotron-emitting electrons does not solve the problem, because the spectrum above the peak is extremely soft: spectral fits require \(\beta<-3\), much steeper than expected from a standard non-thermal electron distribution. A synchrotron self-Compton interpretation is also disfavored, because it would require unrealistically weak magnetic fields \(B'\sim10^{-1}\) mG (see Section \ref{sec:theoretical_interpretation}). We therefore identify external inverse Compton emission as the most natural radiative channel.

In this scenario, freshly heated electrons behind the forward shock cool by upscattering prompt MeV photons that stream through the shock from behind. Prompt emission photons with characteristic energy \(E_t\simeq0.5\)~MeV are scattered to \(E_{\rm IC}\simeq50\)~MeV by electrons with Lorentz factors of order \(\gamma_e\sim20\). Efficient cooling requires the inverse Compton cooling time to be shorter than the expansion time, implying a prompt emission luminosity of at least \(L_{\rm GRB}\gtrsim8\times10^{50}\,{\rm erg\,s^{-1}}\). The luminosity of the high-energy component also requires a large number of radiating leptons. If these leptons were supplied only by ordinary swept-up electrons, the required density would be extreme: \(n\gtrsim10^{7}\,{\rm cm^{-3}}\) for a homogeneous medium or \(A\gtrsim2\times10^{15}\,{\rm g\,cm^{-1}}\) for a wind-like medium with \(\rho=A R^{-2}\).

A natural way to obtain reasonable lepton density without invoking an extreme baryon density is the pair enrichment of the circumburst medium by the prompt radiation front \citep{Madau2000,Meszaros2001,Beloborodov2002}. Prompt emission photons propagate ahead of the blast wave and scatter off ambient electrons. Some of the scattered photons then collide with the primary MeV photon beam and convert into electron--positron pairs. The newly created pairs scatter additional prompt emission photons, triggering runaway pair loading. At the same time, the radiation front transfers momentum to the upstream medium and pre-accelerates it before the forward shock arrives. The shock therefore encounters a pair-rich, already moving upstream, rather than a cold, stationary medium \citep{Beloborodov2005}.

\begin{figure*}
\centering
\includegraphics[width=1\linewidth]{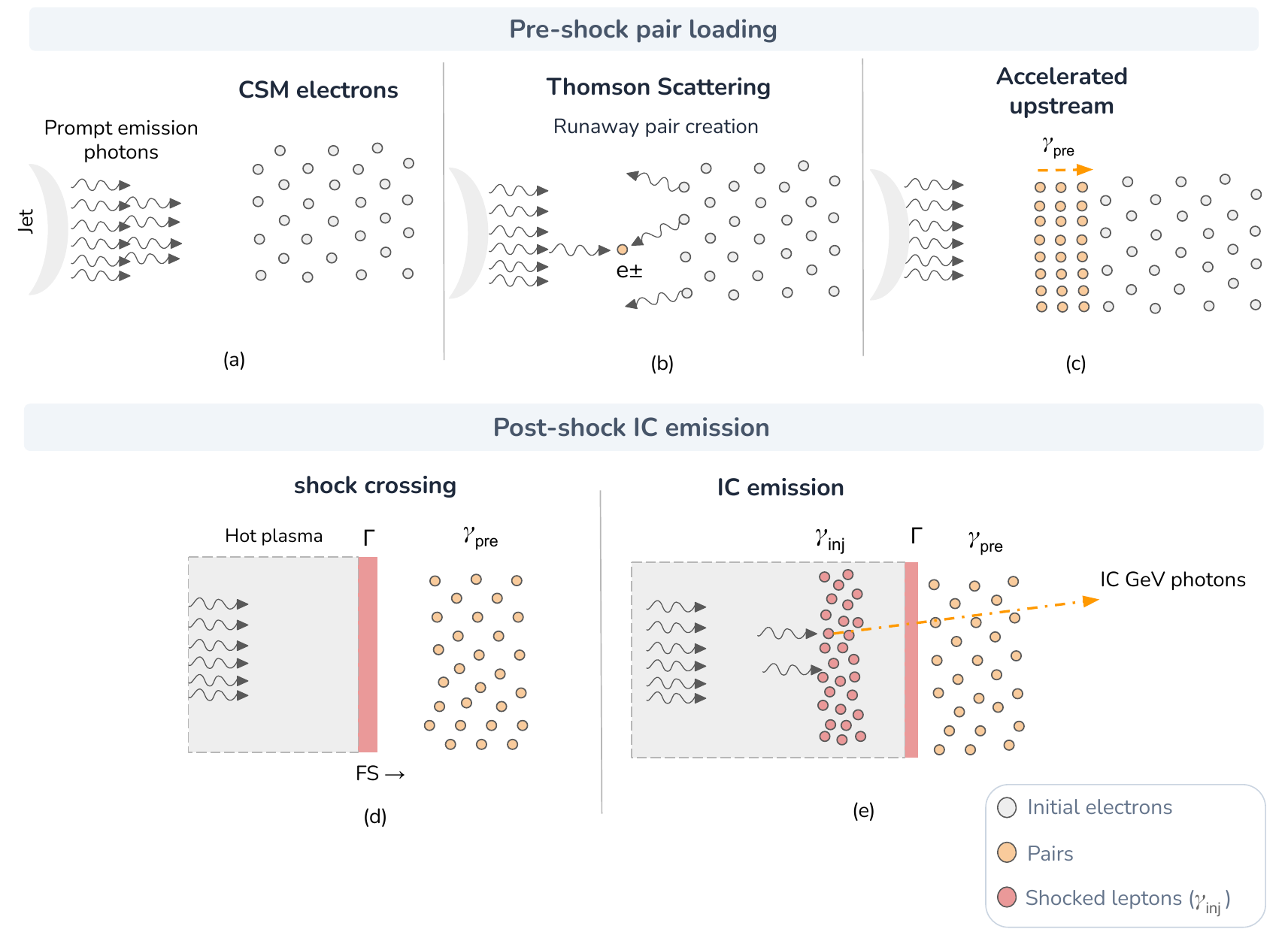}
\caption{
Pair loading, pre-acceleration and external inverse Compton emission in the immediate circumburst medium (model adopted from  \cite{Beloborodov2014}).
(a) Prompt MeV photons from the jet propagate ahead of the blast wave through the circumstellar medium.
(b) A fraction of these photons Thomson-scatters off ambient electrons; the scattered photons then annihilate with the primary MeV beam, triggering runaway creation of electron--positron pairs.
(c) The same radiation front pre-accelerates the pair-enriched upstream medium to Lorentz factor \(\gamma_{\rm pre}\).
(d) The forward shock crosses this modified medium and heats the leptons to $\gamma_{inj}$.
(e) The shocked leptons, with characteristic Lorentz factor \(\gamma_{\rm inj}\), inverse Compton scatter the prompt emission photons. Escaping photons are observed as the fast-fading MeV--GeV afterglow.
}
\label{fig:model_sketch}
\end{figure*}

This modified upstream changes the shock emission in two essential ways \citep{Beloborodov2014}. First, the number of radiating leptons is greatly increased, boosting the inverse Compton luminosity without requiring an unrealistically large baryon density. Second, pre-acceleration reduces the relative Lorentz factor between the shock and the upstream medium, so that the shocked leptons are heated to modest Lorentz factors rather than to the much larger energies expected in a cold upstream. These leptons can then efficiently upscatter prompt MeV photons into the observed MeV--GeV band \citep{Beloborodov2014,Vurm2014,Hascoet2015}. In this model, the fast-fading high-energy afterglow is powered by radiatively efficient external inverse Compton cooling in a pair-loaded stellar wind, as illustrated in Fig.~\ref{fig:model_sketch}.

\begin{figure}[t]
  \centering
  \includegraphics[width=\linewidth]{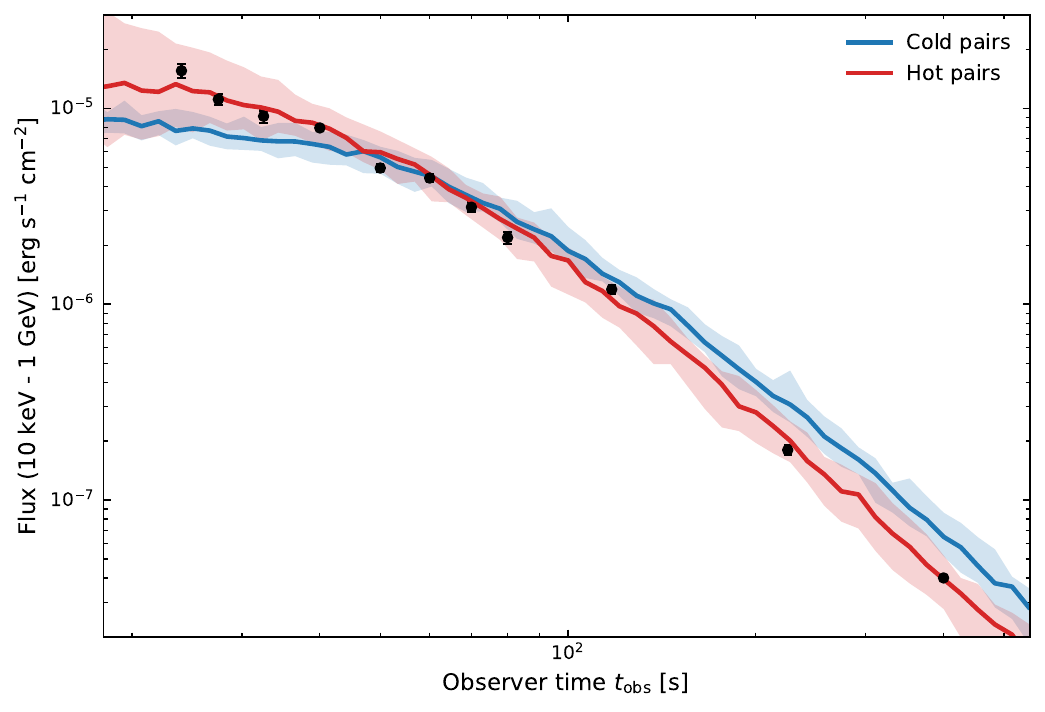}
  \caption{\textbf{Pair-loaded EIC light-curve comparison.}
  Observed 10\,keV--1\,GeV flux of GRB~260226A compared with the light curves of the toy pair-loaded external inverse Compton model.
  The blue band shows cold pair-loading prescription \cite{Beloborodov2005}   and the red
  region shows the adopted hot pairs prescription \cite{Beloborodov2014}.  Solid curves indicate the posterior
  medians and shaded regions show the central 90\% posterior intervals.}
  \label{fig:eic_lightcurve}
\end{figure}

To test this interpretation, we constructed a simplified toy model of inverse Compton emission from a pair-enriched blast wave, described in the Methods. The model follows pair loading and pre-acceleration by the prompt radiation front, computes the heating of swept-up leptons by the forward shock, and includes anisotropic inverse Compton scattering, and photon--photon absorption. We fit the observed bolometric \(10\,{\rm keV}\)--\(1\,{\rm GeV}\) light curve and the spectral peak using a wind-like density profile, \(\rho=A R^{-2}\) (Fig.~\ref{fig:eic_lightcurve}). Because the redshift of GRB\,260226A is not known, the redshift and GRB luminosity were treated as free parameters together with the wind-density parameter. The preferred pair-loaded solution gives \(A\simeq10^{12}\,{\rm g\,cm^{-1}}\), much lower than the value required for an ordinary cold upstream, because the radiating lepton population is dominated by pairs rather than by baryon-associated electrons. This model has previously been applied to model GeV light curves of some GRBs \cite{Beloborodov2014,Vurm2014,Hascoet2015}. However, in these cases, the data in the MeV gap were unavailable, implying 2 orders of magnitude lower wind densities and predicting a much higher spectral peak of the External Inverse Compton radiation: $>10$ GeV instead of $15-60$ MeV. 

This toy model reproduces the observed rapid decline of the flux and the \(15\)--\(60\)~MeV spectral peak for a dense, pair-loaded stellar wind. Such a medium is naturally expected around a massive Wolf--Rayet progenitor and provides the baryon reservoir needed for efficient pair loading and inverse Compton emission. GRB\,260226A therefore shows that the earliest high-energy afterglow can be shaped not only by blast-wave dynamics, but also by radiative feedback from the prompt emission on the circumburst medium \citep{Beloborodov2014}. Related signatures may appear in the MeV band, including absorption features in the prompt spectrum \citep{Oganesyan2026} and narrow annihilation-line emission around 10 MeV \citep{Salafia2026}, as observed in GRB\,221009A \citep{Ravasio2024_line}. The spectra of the prompt emission at 23-35 s in our analysis are cutoff power laws with low-energy photon indices of $\approx-1$, suggesting such a cutoff around $\approx 1.2$ MeV in the rest frame \cite{Oganesyan2026} provided the inferred density of \(A\simeq10^{12}\,{\rm g\,cm^{-1}}\) and the pair-multiplication factor of $\sim 10^2$ in the non-relativistic pair-enrichment region. 

The poorly explored \(1\)--\(100\) MeV band therefore provides a direct probe of the immediate environment of GRB progenitors and of the formation of relativistic external shocks. A substantial improvement in the sensitivity of instruments above 1 MeV \citep{e-ASTROGAM,CE} will be essential for using circumburst environments to distinguish GRB progenitor channels and to study the earliest stages of relativistic-shock formation.

\clearpage

\clearpage

\clearpage

\noindent

\newpage
\section{Methods}\label{sec11}

\subsection{LAT data reduction}\label{sec:LATanalysis}
We performed unbinned likelihood analysis of \textit{Fermi}/LAT data for GRB~260226A ($T_0 = 793795080.958$\,s MET) extending to more than 1\,ks, in the energy range 0.1--1\,GeV, using the \textsc{GTBURST}\footnote{\url{https://fermi.gsfc.nasa.gov/ssc/data/analysis/scitools/gtburst.html}} software. We selected a region of interest covering $12^\circ$ around the source location (${\rm R.A.}= 41.93^\circ$, ${\rm Dec.}= 7.73^\circ$, J2000)\cite{2026GCN.43850....1D}. A standard zenith angle cut of $100^\circ$ was applied to remove Earth-limb contamination. We used the \texttt{P8R3\_TRANSIENT020} event class with corresponding instrument response functions, and included the isotropic particle background and 4FGL catalog sources with fixed normalization. The 2$\sigma$ upper limits (assuming a spectral index of -2) and the 1$\sigma$ flux and index errors are reported in Table~\ref{tab:lat_lc}.

\subsection{GBM data reduction}\label{sec:gbm_data_red}

We downloaded the \gbm\ data ($8\ {\rm keV} - 40\ {\rm MeV}$) of GRB 260226A (GBM trigger \texttt{bn260226443}) from the Fermi GBM Burst catalog\footnote{\url{https://heasarc.gsfc.nasa.gov/W3Browse/fermi/fermigbrst.html}}, and performed a standard data reduction using the Fermi science tool \textsc{GTBURST}. In particular, we considered data from the two sodium iodide (NaI, $8-900$ keV) and one bismuth germanate oxide (BGO, $0.3-40$ MeV) detectors with best observational conditions (i.e., lowest separation angles). For this events, we reduced data from NaI 0 and 1, and BGO 0.

The background analysis was performed through \textsc{GTBURST} by using custom background intervals for the polynomial fit. We chose the interval [$-400$ s, $-200$ s] before GBM trigger. According to our background analysis (see Section~\ref{sec:OSV_timebins} for more details), because of \textit{Fermi} satellite moving along its orbit, NaI 0, NaI 1 and BGO 0 separation angles from the source increase, reaching values $> 60^\circ$ after $\sim500$ s from the burst. Hence, we selected the second background interval between [500 s, 550 s]. 

The outputs of the data reduction are the source, background, and weighted response spectral files that are used for the spectral analysis through the Heasarc package \textsc{XSPEC}\footnote{\url{https://heasarc.gsfc.nasa.gov/xanadu/xspec/}} \citep{Arnaud1996}.

\subsection{LLE data reduction}\label{sec:LLE}
We performed spectral analysis of the prompt emission and early afterglow (up to 500 s) in the $30-100$~MeV energy band using the Fermi LAT Low Energy (LLE) technique \cite{Pelassa2010}, which recovers LAT events that are otherwise excluded from standard analyses due to the strongly energy-dependent point spread function of LAT below 100~MeV \cite{Atwood2009}. The LLE technique fills the spectral gap between the Fermi/GBM-BGO detectors and the standard LAT analysis covering the energy range of $30-100$ MeV. LLE events were selected from the TRANSIENT020E (Extended) class (\texttt{evclass}~$=$~64) with additional quality cuts requiring a reconstructed track in the front section of the silicon tracker and a zenith angle less than  $100^{\circ}$, consistent with standard LLE selection  criteria\footnote{\url{https://heasarc.gsfc.nasa.gov/w3browse/fermi/fermille.html}}. Only events with the  TRANSIENT020E bit explicitly set are retained in the analysis. For each  time bin, a detector response matrix was constructed by combining the  effective area from the \texttt{P8R3\_TRANSIENT020E\_V3} instrument response functions for FRONT-converting events  with the energy dispersion parameterized as a King function. The mean  off-axis angle during each time bin was derived directly from the  spacecraft pointing file (FT2) as an exposure-weighted instantaneous  value, rather than a time-averaged estimate from the livetime cube, which  can be biased toward large off-axis angles during survey-mode  observations. Background spectra were estimated for each time bin by fitting a second-order polynomial to count rates measured in pre-burst ($-100$ to $-5$~s) and post-burst ($700$--$1000$~s) intervals. The interpolated background at the midpoint of each source bin was written to  a background PHA file.

\subsection{Time bins choice and background estimates}\label{sec:OSV_timebins}
GRB 260226A is one of the brightest gamma-ray burst in the GeV energy range. It is detected by \lat\ as soon as $\sim 14$ s after GBM trigger, and it lasts up to a thousand of seconds later (Fig.~\ref{fig:lat_lc}). From a visual inspection, GBM light curve starts approximately at the same time as the LAT one. It comprises two major pulse-like structures lasting until $\sim 40$ s from trigger. However, the GBM burst duration, estimated in the Fermi GBM Burst catalog through the parameter $T_{90}$, extends up to 184.4 s after the trigger. While GBM emission appears as a faint long-lasting tail after $\sim40$ s, LAT emission is bright (with fluxes larger than $\sim 10^{-8}\ {\rm erg\ s^{-1}\ cm^{-2}}$) and slowly fading (see Fig.~\ref{fig:lat_lc}). 

The presence of a large GeV signal in coincidence with a much fainter MeV one motivated us to extend our spectral analysis at late times, after both the prompt-like pulses and the tail-like emission ending at $\sim185$ s. However, the long-lasting MeV-to-GeV emission of this source could lead to an erroneous selection of background time intervals still affected by source signal, making the overall GBM background polynomial fit unreliable.

For this reason, to accurately retrieve the spectra and background of the long-lasting MeV emission at late times, we employed the modified Orbital Subtraction method (OSV\footnote{\url{https://github.com/LudovicoAlt/py3}}) \citep{De_Santis_2024,Fitzpatrick}. The OSV tool exploits the fact that \textit{Fermi} spacecraft returns to the same position every 15 orbits, and that GBM pointing returns to the same sky coordinates every 30 orbits. In particular, OSV tool models the background over the duration of the burst by averaging the background counts from the source sky coordinates across preceding and subsequent \textit{Fermi} orbits. 

We downloaded the daily GBM CSPEC data of days $25^{th}$, $26^{th}$ and $27^{th}$ of February, 2026, from the Fermi GBM daily data catalog\footnote{\url{https://heasarc.gsfc.nasa.gov/FTP/fermi/data/gbm/daily/}}. We used an average of $\pm14$ and $\pm16$ orbits for our analysis. We chose NaI~0 , NaI~1 and BGO~0 detectors due to their better signal-to-noise ratio and low separation angles from the source, as described in Section~\ref{sec:gbm_data_red}. 

The OSV tool outputs the separation angle, $\theta_{sep}$, for each time bin. It shows that  $\theta_{sep}$ of the best detectors NaI 0 and 1 begin to steadily increase around $\sim 150$~s, reaching $\theta_{sep} \simeq 60^\circ$ after $\sim 500$ s. When $\theta_{sep} > 60^\circ$, the source is not visible anymore by a given detector. Therefore, after 500~s, GRB 260226A is not visible anymore by NaI 0 and 1. Instead, between $150-500$~s source counts are still present, but hardly distinguishable from the background obtained through polynomial fitting (Section~\ref{sec:gbm_data_red}).
Meanwhile, NaI 3 separation angle reaches its minimum between $150-500$~s.

Therefore, we selected 12 time bins between $4-150$ s, according to both GBM and LAT light curves morphology. In these bins, GBM data are reduced through standard polynomial fit of the signal during the background intervals discussed in Section~\ref{sec:gbm_data_red}. Conversely, we selected two time bins between $150-500$ s, extracting the source spectral and simulating the background through the OSV tools for BGO 0 an NaI 0, 1 and 3, the latter being the GBM detector with the lowest separation angle in this time interval. We report our time bin selection in Table~\ref{tab:timebins_bestodels}.
Since the last two time bins are chosen quite away from the trigger, the standard response files provided in \gbm\ burst catalog can not be used. Therefore, we generated custom responses for the relevant time intervals and detectors at the source coordinates (${\rm R.A.}=41.93^\circ$, ${\rm Dec.} =7.73^\circ$, J2000 \cite{2026GCN.43850....1D}) using the official Fermi tool GBM Response Generator\footnote{\url{https://fermi.gsfc.nasa.gov/ssc/data/analysis/gbm/DOCUMENTATION.html}}.

\subsection{Spectral fit routine}\label{sec:spectral_fit_routine}
After the data reduction, we performed a time-resolved spectral analysis of GRB 260226A exploiting a rich dataset consisting of GBM, LLE and LAT data. For the fitting process, we used the Heasarc package \textsc{XSPEC} (version 12.14.0b).

We ignored the energy channels outside $8-900$ keV for the NaI detectors, as well as the $30-40$ keV band in order to avoid the iodine K-edge line at 33.17 keV \citep{Meegan2009}. We selected the energy range $300\ {\rm keV} - 40\ {\rm MeV}$ and $30-100$ MeV for BGO and LLE data, respectively. Given the detection of the LAT photon with highest energy at $\sim 1$~GeV, we selected the energy range $100\ {\rm MeV} - 1\ {\rm GeV}$ for LAT spectra. 

For each spectrum, we included the presence of a cross-calibration constant, \texttt{constant} in \textsc{XSPEC} notation, allowing for a $30\%$ variation for each dataset. Moreover, for each model, we measured the logarithm of the $10\ {\rm keV} - 1\ {\rm GeV}$ bolometric flux, $\log F_{\rm bol}$, using the convolutional model \texttt{cflux} in \textsc{XSPEC}. We applied Poisson-Gaussian statistics (\texttt{pgstat}) to standard GBM data and Cash statistics (\texttt{cstat}) \cite{Cash1979} to LLE, LAT and GBM-OSV data.

Initially, we investigated the shape of each time-resolved spectrum by fitting the phenomenological Band function (\texttt{grbm} in \textsc{XSPEC} notation \cite{Band1993}), composed by two power laws smoothly joined around a peak at energies $E_p$. The power law segment below the spectral peak, with spectral index $\alpha$, was adequately describing the low energy dataset of each spectrum. 

However, the Band function models the spectrum above the spectral peak as a single power law segment with photon index $\beta$. Fit residuals showed that our dataset is strongly in tension with this description at higher energies in most of the time bins (Table~\ref{tab:timebins_bestodels}). Specifically, spectra in the second and third bins (1 and 2) showed a large flux suppression at high energies. For these bins, we introduced a high energy exponential cutoff, \texttt{highecut} in \textsc{XSPEC} notation. The overall model (hereafter {BandCutoff}) includes two new parameters: the energy at which the cutoff starts to modify the base spectrum, $E_{\rm start}$, and the energy which regulates the sharpness of the decay, $E_{\rm fold}$. The sum of these two quantities provides the energy at which the flux drops by a factor $1/e$, namely $E_{\rm cutoff}=E_{\rm start} + E_{\rm fold}$. We test different $E_{\rm start}$ values and find consistent results on $E_{\rm cutoff}$ , implying that the position of the $E_{\rm start}$ energy does not affect the main results of the analysis. Therefore, we fix $ E_{\rm start}= 5\ {\rm MeV}$ in order to minimize the number of free parameters.

Conversely, spectra in the following three time bins (3, 4, and 5) showed residuals at high energies both below and above the Band best-fit model, mimicking the presence of a secondary ``bump''. To account for these discrepancies, we introduced in the model the presence of a second component, described as a power law with an exponential cutoff, \texttt{cutoffpl} in \textsc{XSPEC} notation. This model component (hereafter CPL) is added to the primary one, forming a double components model. When a Band function is used to describe the primary continuum, we refer to the overall model as {Band+CPL}. 

We also tested the hypothesis of a primary continuum described by a cutoff power law. This other double components model (hereafter {CPL+CPL}) differs from the Band+CPL model on the description of the region in between the two spectral peaks. In CPL+CPL model, the primary component (the one with spectral peak at lower energies) fades exponentially after the spectral peak, predicting a smaller flux in that spectral region with respect to a Band function. In this scenario, all the emission between the two spectral peaks is provided by the second component (the one with spectral peak at higher energies). Conversely, in the Band+CPL model, the emission between the spectral peaks has to account for the presence of two power law segments (the one above the spectral peak of the first component, and the one below the spectral peak of the second one), sensibly modifying the shape of the secondary spectrum. In both these double components models, we measured $\log {F}_{bol}$ of each component in the $10\ {\rm keV} - 1\ {\rm GeV}$ through the XSPEC convolutional model \texttt{cflux}.

In the remaining time bins (from 6 to 13), the Band function fit returned satisfactory results. However, we noticed than the best-fit estimates of the high energy photon index were particularly soft (i.e., $\beta < -3 $). Therefore, we decided to model these spectra with the simpler CPL model. The list of tested models for each time bin is shown in Table~\ref{tab:timebins_bestodels}. It is worth mentioning that, although the inclusion of cross-calibration constants and the LLE datasets provide a better management of systematics and overall a better description of the spectrum, the results we report in this work are not driven by them. In Fig.~\ref{fig:spectral_analysis_comparison} we show how the main results of our spectral analysis remain consistent when one of both of these features are removed from the analysis.

\subsection{Parameter estimation and spectral models comparison}\label{sec:model_comp_param_estim}
After running the fit of each model for each time bin (Table~\ref{tab:timebins_bestodels}), we produced marginalized posterior distributions of the spectral parameters using a MCMC through the {\sc XSPEC} command {\tt chain}. This analysis returns, for each model parameter, a chain of parameter values whose density gives the parameter probability distribution. We employ the Goodman-Weare algorithm, requiring $N_{walk}=15$  walkers and $N_{iter} = 5\times 10^5$ iterations, for a total of $N_{iter} \times N_{walk} = 7.5 \times 10^6$ samples. Since the starting parameters are far from convergence, we ignore the first $N_{burn} = 10^4 $ steps. The walkers are initialized by drawing from a multi-Normal distribution whose variance matrix is based on the covariance matrix obtained from the previous \textsc{XSPEC} fit. 

From the posterior distribution of the parameters obtained from the spectral fit, we derived parameter distributions of also other quantities of interest. For all the models, we evaluated the bolometric flux as $F_{\rm bol} = 10^{\log F_{\rm bol}}$. In the case of Band and CPL model components, we evaluated the peak energy defined as $E_p = (2+\alpha) \cdot E_c$, where $\alpha$ is the low-energy photon index and $E_c$ is the Band characteristic energy or the CPL cutoff energy, respectively\footnote{In XSPEC, the low-energy photon index of the \texttt{grbm} model is negatively defined, while the one of the \texttt{cutoffpl} model is positively defined. Throughout this paper, we will refer to both using the Band function notation.}. In the case of the BandCutoff model, we evaluated the cutoff energy $E_{cutoff} = E_{start} + E_{fold}$, with $E_{start}$ fixed at 5 MeV. For each relevant parameter, we defined the best-fit value as the median on the posterior distribution, and lower and upper errors were derived from the 16th and 84th percentile of the posterior distribution, respectively.

For each fit, we estimated the value of the relative statistics, $stat$, which is connected to the mixed-likelihood,  $\mathcal{L}$, through the relation $stat = -2\ln \mathcal{L}$. During each fitting procedure, we searched for the parameter values which maximize $\mathcal{L}$, thus minimizing $stat$. To perform model comparison, we used the Akaike Information Criterion (AIC, \cite{Akaike1974}). For each fit in each time bin, we computed ${\rm AIC} = 2k - 2\ln \mathcal{L} = 2k + stat$, where $k = k_0+N_{det} -1$ is the number of free parameters in the fitted model, $k_0$ is the number of free parameters of a given model without calibration constants and $N_{det}$ is the number of detectors employed in the spectral fit. For time bins $0-11$, $N_{det}=5$, while in the last two bins (12 and 13) $N_{det}=6$.

To assess which is the best fit model between the null model 1 and the more complex model 2 (with $k_2 > k_1$), we computed $\Delta {\rm AIC} = {\rm AIC}_1 - {\rm AIC}_2$. We considered a more complex model as being statistically preferred over a simpler one whenever $\Delta {\rm AIC} > 4$ \cite{Burnham2004}. This corresponds to a scenario where the simpler model is $\sim 0.13$ times as probable as the more complex one to minimize the information loss. If  $\Delta {\rm AIC} < 4$, we consider the simpler model as the best-fit. 

We performed the AIC analysis for each temporal bin of the time-resolved spectral analysis where more than one model was tested (time bins $0-5$). Results of the model comparison are reported in Table~\ref{tab:timebins_bestodels}.

\subsubsection{The presence of two spectral components}\label{sec:comparison_2nd_comp}

Model comparison and fit results reveal a clear spectral evolution (Table~\ref{tab:timebins_bestodels} and Fig.~\ref{fig:time_res_individual_spectra}). During the ``prompt'' phase, while the MeV emission reaches its maximum flux and the GeV emission starts to rise, the spectrum is well described by a Band function. In time bins 1 and 2, the spectrum shows $E_p \simeq 600$ keV and $E_{cutoff} \simeq 50$ MeV.

Later, during the ``transition'' phase (time bins 3, 4, and 5), the spectrum is best described by two components: one peaking at $\sim 500$ keV and the other at tens of MeV. In the remaining bins, corresponding to the ``afterglow'' phase, the spectrum is again characterized by a single component peaking at tens of MeV.

These results suggest that the spectrum of GRB 260226A is composed of two main components. One dominates the energy range observed by \gbm, while the other is mainly detected in the \lat\ range. The evolution of these components follows the behavior of the corresponding light curves: the first component dominates while the GBM emission reaches its peak ($t<23$ s), whereas the second component becomes dominant after the GBM emission fades and the LAT emission persists ($t>35$ s). Between these intervals ($23-35$ s), both components contribute significantly to the spectrum.

During this ``transition'' phase, we tested three spectral models: Band, CPL+CPL, and Band+CPL (Section~\ref{sec:model_comp_param_estim}). The Akaike Information Criterion (AIC) strongly favors the presence of two distinct spectral components, with $\Delta{\rm AIC} > 40$ in all three bins. Among the two-component models, the Band+CPL representation is statistically preferred over CPL+CPL, with $\Delta{\rm AIC}_3 = 7.16$, $\Delta{\rm AIC}_4 = 9.64$ and $\Delta{\rm AIC}_5 = 7.58$.

As discussed in Section~\ref{sec:spectral_fit_routine}, the main differences between the CPL+CPL and Band+CPL models (Fig.~\ref{fig:Band/CPL+CPL_comparison}) are the following:
\begin{itemize}
\item[1.] In the Band+CPL model, the first component includes a power-law tail above its spectral peak, unlike the exponential suppression of the CPL+CPL model. As a consequence, the spectral region between the two peaks receives contributions from both components, rather than from the second component alone.\

\item[2.] In the Band+CPL scenario, the emission around $\sim 1$ GeV can still be significantly affected by the first component, whereas in the CPL+CPL model this energy range is entirely dominated by the second component.\\

\end{itemize}

Fig.~\ref{fig:Band/CPL+CPL_comparison} shows the evolution of the total models and their individual components during time bins 3, 4, and 5. Table~\ref{tab:spectral_timeres_2comp_comparison} reports the best-fit parameters and corresponding 1$\sigma$ uncertainties for both spectral descriptions.

Despite the subtle difference between the two models, namely the exponential cutoff versus the power-law tail above the first component peak, they imply very different phenomenological interpretations. In the CPL+CPL scenario, the two components exhibit little spectral evolution during the transition phase (Fig.~\ref{fig:Band/CPL+CPL_comparison}, left column). At later times ($35-500$ s) the spectrum maintains a shape and temporal evolution consistent with the second component observed during the transition phase (see Fig.~\ref{fig:time_res_individual_spectra}). However, the first component appears somewhat inconsistent with the prompt phase, being suppressed at lower energies than in the previous bins.

Conversely, in the Band+CPL scenario, the first component evolves more smoothly from the ``prompt'' to the ``transition'' phase. However, the second component undergoes a strong hardening, with the photon index evolving from $\alpha \simeq-1.2$ to $\alpha \simeq+0.5$ (Table~\ref{tab:spectral_timeres_2comp_comparison}). In the following bins ($t>35$ s), the spectrum shows a similar $E_p$, but the photon index becomes significantly softer again ($\alpha \simeq-1.5$).

Overall, our results support the presence of two main spectral components emitting in different energy bands: one peaking at hundreds of keV in the GBM range, and the other at tens of MeV in the LAT range. Their temporal evolution defines three distinct phases. During the first phase (bins 0, 1, and 2), the GBM component dominates and displays properties consistent with standard prompt emission. During the last phase (bins 6 to 13), the LAT component dominates after the prompt emission fades at $\sim35$~s, and is therefore associated with the afterglow. During the transition phase, both components contribute simultaneously in different spectral regions.

Although the statistically preferred Band+CPL model describes a smoother evolution of the prompt component, it requires the second component to evolve in a way that differs from the afterglow observed at later times, becoming initially narrow and hard before softening again. In contrast, the CPL+CPL model describes the rise of a secondary component that is spectrally consistent with the emission later identified as afterglow, and that gradually dominates as the prompt component fades. For this reason, and for additional theoretical motivations discussed in
Sec. \ref{sec:theoretical_interpretation}, we favor the CPL+CPL interpretation despite the outcome of the statistical model comparison.

\subsection{Temporal analysis}\label{sec:temporal_analysis}
To characterize the temporal evolution of the second spectral component, we modeled its bolometric flux profile with an empirical broken power law. We adopted a Bayesian framework with uniform priors on the temporal slopes, $a_1$ and $a_2$, and log-uniform priors on the break time and normalization, $t_b$ and $F_{\rm bol}^0$, respectively. Specifically, we adopted the following prior ranges: $\log \left( F_{\rm bol}^0 [{\rm cgs}]\right) \in (-5, 1)$, $\log(t_b/{\rm s}) \in (1.15, 2.70)$, $a_1 \in (-4, 0)$ and $a_2 \in (-6, 0)$. The posterior distribution was sampled using a Markov chain Monte Carlo method implemented in the \texttt{emcee} package \citep{2013PASP..125..306F}. Convergence was assessed via autocorrelation time, requiring a chain length that is more than 100 times the estimated autocorrelation time and stability of this estimate within 1\%. The corner plots of the posterior distributions are shown in Fig.~\ref{fig:Fbol_posterior}.

It is worth mentioning that the reference time of the second spectral component, $T^{ag}_0$, does not necessarily coincide with the GRB trigger time, possibly affecting the temporal decay indices inferred from the analysis. To test this possibility, we repeated the fit to the bolometric flux of the second component by adopting $T^{ag}_0 = 20$~s, corresponding to approximately 3~s before the peak of the high-energy component. The resulting temporal evolution is well described by a broken power law, with an initial decay slope of $a_1 \sim -0.6$ that steepens to $a_2 \sim -2.2$ approximately 45~s after the adopted $T^{ag}_0$. This demonstrates that the presence of a steep decay phase is robust against the choice of $T^{ag}_0$.

\clearpage
\subsection{Theoretical interpretation}\label{sec:theoretical_interpretation}

We investigate the physical origin of the fast high-energy afterglow of GRB~260226A.  The observed high-energy (HE) component has the following properties.  First, its onset occurs at $t_{\rm obs}\simeq20\,{\rm s}$ after the trigger.  Second, the time-resolved HE spectrum is well described by a cutoff power law with a photon index $\Gamma_{\rm ph}\simeq1.5$ below the cutoff and a peak in $\nu F_{\nu}$ at $E_{\rm p}\simeq 15$--$60\,{\rm MeV}$.  Third, the spectrum above the peak is extremely soft: independent Band-function fits require a high-energy spectral index $\beta<-3$.  Fourth, the bolometric HE flux (10 keV - 1 GeV) initially decays approximately as $t^{-1.5}$ and steepens to approximately $t^{-3}$ after $t_{\rm obs}\simeq65\,{\rm s}$.  Fifth, the HE fluence integrated over $20$--$500\,{\rm s}$  is $4.7 \times 10^{-4} \, \rm erg \, cm^{-2}$ is comparable to the fluence of the prompt MeV emission $\approx 3.5 \times 10^{-4} \, \rm erg \, cm^{-2}$ (0-35 s), implying an upper limit for the prompt emission efficiency of 0.4.  Finally, the HE bolometric flux of the afterglow is comparable to that of the prompt MeV radiation at the simultaneous time-bins and it is order of magnitude lower compared to the brightest prompt emission time bins.  Although the measured HE peak energy of the afterglow is not strictly constant, its variations are modest compared with the dynamical range in flux. Therefore in our estimates below we use $E_{\rm p}\simeq50\,{\rm MeV}$ as a representative value.

The HE afterglow light curve already disfavors a standard adiabatic external-shock afterglow. For a usual relativistic adiabatic blast wave one expects a much shallower decline, of order $t^{-1}$.  Even the fully radiative Blandford--McKee limit gives a decay close to $t^{-10/7}$, which may account for the first $\sim65\,{\rm s}$ but not for the later $t^{-3}$ decline.  Below, we first examine which radiative processes can produce the HE photons, before constructing a more specific dynamical model. 

\subsubsection{Inconsistency with a synchrotron origin}

Suppose first that the $50\,{\rm MeV}$ photons are synchrotron photons emitted by electrons of Lorentz factor $\gamma_e$ in a comoving magnetic field $B'$ and a source moving with bulk Lorentz factor $\Gamma$.  Ignoring cosmological redshift, we infer
\begin{equation}
  B' \simeq 2.9\times10^5\,
  E_{\rm syn,50}\,\gamma_{e,4}^{-2}\,\Gamma_2^{-1}\,{\rm G},
  \label{eq:Bsynreq}
\end{equation}
where $E_{\rm syn,50}=E_{\rm syn}/50\,{\rm MeV}$, $\gamma_{e,4}=\gamma_e/10^4$ and $\Gamma_2=\Gamma/100$.

If these photons are produced by the characteristic electrons accelerated at the forward shock, then for a power-law distribution $dN_e/d\gamma_e\propto\gamma_e^{-p}$ with $p=3$ (to account for steep spectra at $>50$ MeV) one has
\begin{equation}
  \gamma_m \simeq 9.2\times10^3\,\epsilon_{e,-1}\Gamma_2 .
  \label{eq:gammam_synch}
\end{equation}
Equation~(\ref{eq:Bsynreq}) then requires
\begin{equation}
  B'\simeq 3.4\times10^5\,
  E_{\rm syn,50}\,\epsilon_{e,-1}^{-2}\Gamma_2^{-3}\,{\rm G} .
\end{equation}
Such high magnetic fields would indeed allow the electrons to be deep in the fast-cooling regime with the comoving synchrotron cooling time 
\begin{equation}
  t'_{\rm syn}
  \simeq 7\times10^{-7}\,
  E_{\rm syn,50}^{-2}\epsilon_{e,-1}^{3}\Gamma_2^5\,{\rm s} .
\end{equation}
However, shock-generated magnetic fields are expected to be many orders of magnitude weaker.  For a homogeneous external medium,
\begin{equation}
  B'
  \simeq 0.4\,n_0^{1/2}\epsilon_{B,-4}^{1/2}\Gamma_2\,{\rm G},
\end{equation}
where $n_0=n/(1\,{\rm cm^{-3}})$.  For a wind medium, $\rho=A R^{-2}$ and $R\simeq2c t_{\rm obs}\Gamma^2/(1+z)$ give
\begin{equation}
  B'\simeq 18\,A_{11.7}^{1/2}\epsilon_{B,-4}^{1/2}
  \Gamma_2^{-1}t_{{\rm obs},1.3}^{-1}(1+z)\,{\rm G},
\end{equation}
where $A_{11.7}=A/(5\times10^{11}\,{\rm g\,cm^{-1}})$ and $t_{{\rm obs},1.3}=t_{\rm obs}/20\,{\rm s}$.  This discrepancy excludes synchrotron emission from the characteristic forward shock accelerated electrons.

Instead, one could associate the $50\,{\rm MeV}$ photons with the highest-energy electrons that cool within a dynamical time.  This does not solve the spectral problem.  If the observed peak is the synchrotron cooling break, the observed low-energy photon index $\Gamma_{\rm ph}\simeq1.5$ requires $p =2$.  The spectrum $> 50 $ MeV would then reflect the injected non-thermal distribution.  For a standard $p\simeq2$ distribution one expects $\Gamma_{\rm ph}\simeq p/2+1\simeq2$, or equivalently a Band high-energy index $\beta\simeq-2$, not the observed $\beta<-3$. Thus a standard synchrotron origin cannot explain the HE afterglow spectrum.

\subsubsection{Inconsistency with a synchrotron self-Compton origin}

The HE component could alternatively be inverse Compton radiation.  In a synchrotron self-Compton (SSC) scenario, the seed photons are synchrotron photons produced by the same electrons which upscattered them.  To keep the observed $10\,{\rm keV}$--$1\,{\rm GeV}$ band in the first IC generation, the characteristic electrons must be sufficiently energetic that second-order IC scattering of the lowest energy photon observed  is already Klein--Nishina suppressed.  This constraints
\begin{equation}
  \gamma_m \gtrsim {0.4m_ec^2\over E'_{\rm min}}
  \simeq 2\times10^3\,\Gamma_2,
\end{equation}
where $E'_{\rm min}\simeq10\,{\rm keV}/\Gamma$ is the lowest observed photon energy in the comoving frame. The synchrotron seed photons that would be upscattered by these electrons to $10\,{\rm keV}$ have observed energy
\begin{equation}
  E_{\rm syn}
  \simeq 2.5\times10^{-3}\Gamma_2^{-2}\,{\rm eV}.
\end{equation}
Producing such low-energy synchrotron photons with $\gamma_m\simeq2\times10^3\Gamma_2$ requires
\begin{equation}
  B'\sim 4\times10^{-4}\,\Gamma_2^{-5}\,{\rm G},
\end{equation}
far below the already weak magnetic fields expected behind a relativistic external shock.  We therefore do not consider SSC a natural explanation of the observed bright, rapidly fading HE component.

\subsubsection{External inverse Compton scenario}

The remaining possibility is external inverse Compton (EIC) emission, in which shocked electrons in the reverse/forward shock cool by prompt MeV photons that come from behind \citep{Beloborodov2005b,Wang2006,Fan2006,Fan2008,Kumar2014,Kimura2019,Zhang2020}.  For prompt photons with characteristic observed energy $E_t\simeq0.5\,{\rm MeV}$, the IC photon energy is of order
\begin{equation}
  E_{\rm IC}\sim {1\over3}\gamma_e^2 E_t.
\end{equation}
Hence $E_{\rm IC}\simeq50\,{\rm MeV}$ requires
\begin{equation}
  \gamma_e\sim\left({3E_{\rm IC}\over E_t}\right)^{1/2}\simeq17,
\end{equation}
consistent with mildly relativistic electrons.  Larger electron Lorentz factors are allowed if the intrinsic IC spectrum extends above $50\,{\rm MeV}$ and photon--photon absorption shapes the observed cutoff.

Fast EIC cooling requires $t'_{\rm IC}<t'_{\rm exp}$, where
\begin{equation}
  t'_{\rm IC}={3m_ec\over4\sigma_T\gamma_e U'},
  \qquad
  t'_{\rm exp}={R\over \Gamma c}.
\end{equation}
For a radially beamed prompt emission photon field moving in the same direction as the blast wave,
\begin{equation}
  U'\simeq {U\over4\Gamma^2},
  \qquad
  U={L_{\rm GRB}\over4\pi R^2c}.
\end{equation}
Using $R\simeq2ct_{\rm obs}\Gamma^2/(1+z)$ gives
\begin{equation}
  L_{\rm GRB}\gtrsim
  8.3\times10^{50}\,
  {\Gamma_2^5 t_{{\rm obs},1.3}\over \gamma_{e,1.3}(1+z)}
  \,{\rm erg\,s^{-1}},
  \label{eq:eic_fast_cooling}
\end{equation}
where $\gamma_{e,1.3}=\gamma_e/20$.  

A second constraint follows from the luminosity.  In the fast-cooling limit, the EIC luminosity is approximately the luminosity carried by relativistic electrons accelerated at the shock,
\begin{equation}
  L_e\simeq \epsilon_e L_{\rm sh},
  \qquad
  L_{\rm sh}\sim 4\pi R^2\rho_{\rm ext}\Gamma^2c^3 .
\end{equation}
For a homogeneous medium, $\rho_{\rm ext}=n m_p$, and therefore
\begin{equation}
  n\simeq1.8\times10^6\,
  L_{e,50}\epsilon_{e,-1}^{-1}\Gamma_2^{-2}R_{16}^{-2}
  \,{\rm cm^{-3}} .
  \label{eq:n_req_generic}
\end{equation}
If we further require $L_e\gtrsim L_{\rm GRB,min}$ from Eq.~(\ref{eq:eic_fast_cooling}), this becomes
\begin{equation}
  n\gtrsim1.5\times10^7\,
  {\Gamma_2^3 t_{{\rm obs},1.3}\over\gamma_{e,1.3}(1+z)}
  \epsilon_{e,-1}^{-1}R_{16}^{-2}
  \,{\rm cm^{-3}} .
  \label{eq:n_req_fastcool}
\end{equation}
Equivalently, after substituting $R\simeq2ct_{\rm obs}\Gamma^2/(1+z)$,
\begin{equation}
  n\gtrsim10^7\,
  {1+z\over \gamma_{e,1.3}\,\Gamma_2\,t_{{\rm obs},1.3}}
  \epsilon_{e,-1}^{-1}
  \,{\rm cm^{-3}} .
\end{equation}
For a wind medium, $\rho_{\rm ext}=A R^{-2}$, and
\begin{equation}
  A\simeq2.9\times10^{14}\,
  L_{e,50}\epsilon_{e,-1}^{-1}\Gamma_2^{-2}
  \,{\rm g\,cm^{-1}} .
  \label{eq:A_req_generic}
\end{equation}
Combining this with Eq.~(\ref{eq:eic_fast_cooling}) gives
\begin{equation}
  A\gtrsim2.4\times10^{15}\,
  {\Gamma_2^3 t_{{\rm obs},1.3}\over\gamma_{e,1.3}(1+z)}
  \epsilon_{e,-1}^{-1}
  \,{\rm g\,cm^{-1}} .
  \label{eq:A_req_fastcool}
\end{equation}
Thus, if the observed HE luminosity is supplied by ordinary swept-up electrons, densities much larger than those of a typical interstellar medium or Wolf--Rayet wind are required.

The associated Thomson optical depth for GRB X-ray photons is
\begin{equation}
  \tau_T
  \simeq1.2\times10^{-2}\,
  L_{e,50}\epsilon_{e,-1}^{-1}\Gamma_2^{-2}R_{16}^{-1},
  \label{eq:tauT_req_generic}
\end{equation}
for both the homogeneous and the wind circumburst medium.  Using the Eq.~(\ref{eq:eic_fast_cooling}), we get
\begin{equation}
  \tau_T\gtrsim8.3\times10^{-2}\,
  \Gamma_2\,\gamma_{e,1.3}^{-1}\epsilon_{e,-1}^{-1} .
  \label{eq:tauT_req_fastcool}
\end{equation}
For a GRB luminosity above the lower limit, the Thomson depth approaches unity.  If the upstream is itself moving outward with Lorentz factor $\gamma_{\rm pre}$, the optical depth seen by an outward photon is significantly reduced \citep{Abramowicz1991}. Thus, a pre-accelerated upstream would resolve the problem of opacity. 

\subsubsection{External inverse Compton emission in a pair-enriched medium}

A natural way to obtain a large lepton density without invoking an extreme baryon density is to pair-load and pre-accelerate the circumburst medium by the prior propagation of the prompt radiation front \citep{Madau2000,Beloborodov2002}. A small fraction of prompt photons scatter on circumburst electrons and the scattered photons then annihilate with the primary collimated MeV photons and convert to $e^\pm$ pairs.  Since the newly created pairs also scatter prompt emission photons, this leads to runaway pair creation. The same radiation front transfers momentum to the external medium and pre-accelerates it. These effects are controlled by the radiation-front compactness \citep{Beloborodov2002}
\begin{equation}
  \xi(R)={\sigma_T E_{\rm GRB,ahead}(R)\over4\pi R^2m_ec^2},
\end{equation}
where $E_{\rm GRB,ahead}$ is the prompt emission energy that has overtaken the forward shock.  In the analytic cold-front approximation, the resulting $Z_\pm(\xi)$ (pair multiplicity) and $\gamma_{\rm pre}(\xi)$ are computed in \cite{Beloborodov2002,Beloborodov2005}.  The more detailed calculations show that the profile depends on the prompt emission spectrum and that more energetic radiation fronts can produce larger pair multiplicities at fixed $\gamma_{\rm pre}$ \citep{Beloborodov2014}.

The forward shock heats the pair-loaded upstream to a characteristic lepton Lorentz factor \citep{Beloborodov2014}
\begin{equation}
  \gamma_{\rm inj}
  =\Gamma_{\rm rel}\left(\gamma_{\rm th}
  +\epsilon_e{\mu_e m_p\over Z_\pm m_e}\right),
  \label{eq:gamma_inj_model}
\end{equation}
where $\Gamma_{\rm rel}=\Gamma\gamma_{\rm pre}(1-\beta_\Gamma\beta_{\rm pre})$ is the relative Lorentz factor between the shocked fluid and the upstream, $\gamma_{\rm th}=1$ in the cold-lepton approximation, $\mu_e=1$ for hydrogen, and $\epsilon_e$ controls ion-to-lepton energy transfer. When $Z_\pm\gg m_p/m_e$, most of the dissipated energy is already carried by leptons and the baryon term is suppressed.  The key point is that pre-acceleration lowers $\Gamma_{\rm rel}$, while pair loading increases the number of radiating leptons.  The combination naturally gives large luminosity and modest $\gamma_{\rm inj}$, placing the IC peak in the MeV--GeV range.

\subsubsection{Toy EIC model in the pair-enriched medium}\label{sec:toy_eic_pair}

We modelled the high-energy component with a simplified external inverse Compton
(EIC) calculation in a pair-loaded blast wave.  The model follows the physical
picture of refs.~\citep{Beloborodov2014,Vurm2014}, but it is not a full
radiation transfer calculation. The purpose is to test whether the
observed 10\,keV--1\,GeV light curve and the range of the observed
$\nu F_\nu$ peak between 15-60 MeV can be reproduced with the pair enrichment, pre-acceleration
and photon--photon absorption expected when the prompt radiation front overtakes
the circumburst medium.

The prompt emission is described by an isotropic-equivalent luminosity
$L_{\rm GRB}$ lasting for $T_{\rm obs}=35\,{\rm s}$.  We use a Band spectrum
with observed peak energy $E_{p, {\rm obs}}=0.5\,{\rm MeV}$ and photon indices
$\alpha=-1$ and $\beta=-2.5$.  For a given redshift $z$, the rest frame peak
energy is $E_{p}=(1+z)E_{p, {\rm obs}}$ .
The blast wave propagates in a wind medium  $\rho(R)=A R^{-2}$,
where $A$ is one of the fitted parameters.  We use the radiation-front coordinate
\begin{equation}
  \ell(R)=\int_0^R {dR'\over 2\Gamma_{\rm FS}^2(R')},
  \label{eq:ell_coordinate_methods}
\end{equation}
where $\Gamma_{\rm FS}$ is the Lorentz factor of the forward shock.  The finite
prompt emission duration corresponds to the rest frame width
$cT_{\rm src}=cT_{\rm obs}/(1+z)$.  Prompt emission seed for the EIC radiation is included only
when $\ell(R)<cT_{\rm src}$ .
The decoupling radius $R_1$ is defined by $\ell(R_1)=cT_{\rm src}$,after this
point the prompt photons have overtaken the forward shock.  We join the solution smoothly to the adiabatic wind scaling
$\Gamma\propto R^{-1/2}$ for $R>R_1$.

At each radius the upstream pair loading and pre-acceleration are controlled by
the dimensionless parameter
\begin{equation}
  \xi(R)={\sigma_{\rm T} E_{\rm GRB}(<\ell/c)\over4\pi R^2m_ec^2},
  \label{eq:xi_methods}
\end{equation}
where for the prompt emission light curve is approximated to have a constant luminosity of a duration $T_{\rm src}$,
$E_{\rm GRB}(<\ell/c)=L_{\rm GRB}\ \min(\ell/c,T_{\rm src})$.
We considered two prescriptions for $Z_\pm(\xi)$ and
$\gamma_{\rm pre}(\xi)$.  The firstuses the analytic cold-front
approximation \cite{Beloborodov2005}. The second uses a
hot pairfront profile guided by the numerical calculations \cite{Beloborodov2014}, appropriate for a prompt spectrum peaking around
MeV energies. In the second case, the profile gives larger $Z_\pm$ at fixed
$\gamma_{\rm pre}$ than the cold approximation.  We use these two profiles as limiting descriptions.

The blast wave Lorentz factor is computed from the pressure balance
approximation between the forward and reverse shocks \citep{Beloborodov2014}.  We use
\begin{equation}
  \Gamma\simeq {\Gamma_{\rm ej}\over
  \left[1+2\Gamma_{\rm ej}^2
  \left({4\pi A c^3(1+Z_\pm m_e/\mu_e m_p)
  \over L_{\rm ej}\gamma_{\rm pre}(1+\beta_{\rm pre})}\right)^{1/2}\right]^{1/2}},
  \label{eq:Gamma_pressure_balance_methods}
\end{equation}
where $\beta_{\rm pre}=(1-\gamma_{\rm pre}^{-2})^{1/2}$. We adopt $\mu_e=2$, appropriate for a hydrogen poor Wolf-Rayet wind dominated by helium, consistent with the massive star progenitor scenario discussed above. We set $L_{\rm ej}=1.5\ L_{\rm GRB}$, corresponding to a
prompt radiative efficiency of approximately 0.4, and use
$\Gamma_{\rm ej}=10^3$.  The result is insensitive to the precise value of
$\Gamma_{\rm ej}$ as long as the reverse shock is relativistic.

The relative Lorentz factor between the shocked blast wave and the
pre-accelerated upstream is
\begin{equation}
  \Gamma_{\rm rel}=\Gamma\gamma_{\rm pre}(1-\beta\beta_{\rm pre}).
  \label{eq:Gamma_rel_methods}
\end{equation}
The shocked lepton Lorentz factor is then \citep{Beloborodov2014}
\begin{equation}
  \gamma_{\rm inj}=\Gamma_{\rm rel}
  \left(\gamma_{\rm th}+\epsilon_e{\mu_e m_p\over Z_\pm m_e}\right).
  \label{eq:gamma_inj_methods}
\end{equation}
In the fits shown here we set $\epsilon_e=0$ and $\gamma_{\rm th}=1$, so to probe only the pairs component.

The number of radiating leptons swept by the forward shock in a radial shell is
\begin{equation}
  dN_\ell={4\pi A\over\mu_e m_p}Z_\pm(R)\,dR .
  \label{eq:dN_methods}
\end{equation}
The shell energy available for the EIC radiation is
\begin{equation}
  dE_{\rm cool}=dN_\ell\,\Gamma\,
  \left(\gamma_{\rm inj}-1\right)m_ec^2 f_{\rm cool}.
  \label{eq:shell_energy_methods}
\end{equation}
The cooling fraction is
\begin{equation}
  f_{\rm cool}=1-\exp\left(-{t'_{\rm exp}\over t'_{\rm IC}}\right),
  \qquad
  t'_{\rm exp}={R\over\Gamma c} .
\end{equation}
For a radially beamed prompt emission field, the comoving photon energy density is
\begin{equation}
  U'={L_{\rm GRB}\over16\pi R^2c\Gamma^2} .
\end{equation}
The IC cooling time is evaluated as
\begin{equation}
  t'_{\rm IC}={3m_ec\over4\sigma_{\rm T}U'\gamma_{\rm inj}{\eta}_{\rm KN}},
\end{equation}
where ${\eta}_{\rm KN}\le1$ is Klein--Nishina suppression factor.
Equation~(\ref{eq:shell_energy_methods}) gives only the energy budget, while the
observed light curve is obtained after sampling photon energies, directions,
arrival times and photon-photon absorption.

Target photons are drawn from the photon number distribution of theBand
function.  In the comoving frame of the shock the electrons are isotropic.  For an
electron of speed $\beta_e$, we define $\mu_e=\cos\theta_e$, where $\theta_e$ is the angle between the electron velocity and the incoming
photon beam in the comoving frame.  The scattering probability is
weighted by $dP\propto (1-\beta_e\mu_e)\,d\Omega_e$. 
This is the anisotropic EIC effect \citep{Aharonian1981, Brunetti2000}: electrons moving against the incoming beam
scatter more efficiently.  Each event is transformed to the electron
rest frame, a Klein--Nishina scattering angle is sampled, and
the scattered photon is transformed back to the lab frame.  Therefore the same
Monte-Carlo event determines $E_{\rm lab}$, $\mu_{\rm lab}$, $t_{\rm obs}$ and
$\tau_{\gamma\gamma}$.  Packet weights are proportional to the scattered
lab-frame energy and are normalised so that the ensamble of photon packets carries the
shell energy of Eq.~(\ref{eq:shell_energy_methods}).

The observer time of an escaped packet is
\begin{equation}
  t_{\rm obs}=(1+z)\left({\ell\over c}+{R(1-\mu_{\rm lab})\over c}\right),
  \label{eq:tobs_packet_methods}
\end{equation}
where $\mu_{\rm lab}=\cos\theta_{\rm lab}$ and $\theta_{\rm lab}$ is the photon
angle relative to the radial direction in the lab frame.  This expression
includes both the delay of the blast wave behind the leading prompt front and
the angular delay of EIC photons.  The model light curve is obtained by summing
escaped packet energies in observer-time bins and in the observed
10\,keV--1\,GeV band.

Photon-photon absorption is computed against the unscattered prompt radiation, as pointed out to be the dominant source of photon-photon absorption \citep{Beloborodov2014}, using the Breit--Wheeler cross-section.  We do not include a pair cascade in our toy model.

The free parameters of the fit are $\log_{10}(1+z)$,
$\log_{10}L_{\rm GRB}$ and $\log_{10}A$. We adopted the following flat prior ranges: $\log \left( 1+z) \right) \in (4\times10^{-3}, 0.95)$, $\log(L_{\rm iso}) [\rm erg/s] \in (49, 55)$, $\log(A) \, [\rm g \, cm^{-1}] \in (8, 14)$.

We fit the observed 10\,keV--1\,GeV
light curve with the observed $\nu F_\nu$ peak energy range (15-60 MeV) as a prior
using a flat cosmology
with $H_0=70\,{\rm km\,s^{-1}\,Mpc^{-1}}$, $\Omega_m=0.3$ and
$\Omega_\Lambda=0.7$. The MCMC was run independently for two pair-enrichment profiles \citep{Beloborodov2002,Beloborodov2014}.  We show the resulting posterior distributions in
Fig.~\ref{fig:eic_posteriors} and the corresponding 90\% posterior model bands
for the light curve in Fig.~\ref{fig:eic_lightcurve}.  The toy model should be
interpreted as a viability test rather than as a unique physical reconstruction:
the full radiation transfer for the observed time-dependent prompt emission spectrum
is not solved.  Nevertheless, the pair-enriched upstream naturally supplies a
large number of radiating leptons and reduces the ordinary density and Thomson-depth
requirements of a non pair-loaded EIC model.

\clearpage

\backmatter

\bibliography{bibliography}

\section*{Declarations}
\bmhead{Author information statement}
Correspondence and requests for materials should be addressed to Biswajit Banerjee, Alessio Mei and Annarita Ierardi.

\bmhead{Acknowledgments} S.M., A.M., S.P. acknowledge support from the INAF grant no. 1.05.24.03.04. A.Mei acknowledges support by funding from the Italian Space Agency, contract ASI/INAF n.I/004/11/6. G.O., A.I. and P.T. acknowledge support from the ASI-GSSI contract n. 2025-5-U.0: “Gamma-ray bursts: a probe of multi-messenger and extreme Astrophysics”. S.R. acknowledges support from the Astrophysics Center for Multi-messenger Studies in Europe (ACME), funded under the European Union’s Horizon Europe Research and Innovation Program, Grant Agreement No. 101131928.
\bmhead{Author contribution statement}

B.B. carried out LAT and LLE data reduction and analysis. A.Mei carried out GBM data reduction and analysis. S.Mo. and G.O. performed GBM analysis through the OSV tool. A.L.D.S. contributed in the adaptation and update of the OSV pipeline for this case study. A.Mei and A.I. performed the joint spectral analysis and model comparison. A.I. analysed the temporal properties of this source. S.M. and G.O. developed the theoretical model used to interpret the afterglow emission. B.B., A.Mei, A.I., S.M. and G.O. lead the paper writing. S.Mo. and E.K. gave significant inputs on data interpretation. All authors contributed to discussions and editing of the paper. B.B. produced Fig. \ref{fig:lat_lc} and \ref{fig:LLE_test}. A.Mei produced Fig. \ref{fig:time_res_spectral_analysis+LC}, \ref{fig:time_res_individual_spectra} and \ref{fig:Band/CPL+CPL_comparison}. A.I. produced Fig. \ref{fig:second_component}, \ref{fig:spectral_analysis_comparison} and \ref{fig:Fbol_posterior}. S.M. produced Fig. \ref{fig:model_sketch}. G.O. produced Fig. \ref{fig:model_sketch} and \ref{fig:eic_posteriors}.

\bmhead{Conflict of interest}
We declare no conflicts of interest.

\bmhead{Data availability}
\lat\ raw data are public and can be downloaded using the software \textsc{gtburst}:\\
\texttt{https://fermi.gsfc.nasa.gov/ssc/data/analysis/scitools/gtburst.html}\\
\lat\ 2nd GRB catalog data are available at the following link:
\texttt{https://www-glast.stanford.edu/pub\textunderscore data/953/}\\
\gbm\ raw data are publicly available at the following link: \texttt{https://heasarc.gsfc.nasa.gov/W3Browse/fermi/fermigbrst.html}\\

\bmhead{Code availability}

\textsc{Heasoft}, \textsc{XSPEC} and \textsc{PyXspec} are freely available online at the following links:\\
\texttt{https://heasarc.gsfc.nasa.gov/docs/software/heasoft/}\\
\texttt{https://heasarc.gsfc.nasa.gov/xanadu/xspec/}\\
\texttt{https://heasarc.gsfc.nasa.gov/docs/xanadu/xspec/python/html/index.html}\\
\textsc{gtburst} is one of the Fermi Science Tools package, freely available at the following link:\\
\texttt{https://fermi.gsfc.nasa.gov/ssc/data/analysis/software/}\\
The details of the \textsc{gtburst} analysis can be found here:\\
\texttt{https://fermi.gsfc.nasa.gov/ssc/data/analysis/scitools/gtburst.html}\\
\textsc{emcee} is a python package, available at the following link:\\
\texttt{https://emcee.readthedocs.io/en/stable/user/install/}\\
All reduced data and scripts are available from the corresponding authors upon reasonable request.\\

\newpage 
{\noindent  \bf Supplementary Materials.}
\setcounter{figure}{0}
\setcounter{table}{0}
\renewcommand{\thefigure}{S\arabic{figure}}
\renewcommand{\thetable}{S\arabic{table}}

\begin{table}[h]
\centering
\caption{LAT likelihood analysis results. All fluxes are estimated in the energy band 0.1-1\,GeV. The photon index and the Test Statistic are presented in the last two columns. We reported the 1$\sigma$ errors for the flux.}
\label{tab:lat_lc}
\setlength{\tabcolsep}{6pt}
\begin{tabular}{l c c c c}
\hline\hline
Bin & ${\rm t-T_{0}}$  &
Flux & $\Gamma_{\rm LAT}$  & TS \\
 & [s] & {($\rm erg\,cm^{-2}\,s^{-1}$)} & & \\
\hline
0 & 14.0--18.0    & $(2.88\pm2.55)\times10^{-8}$ & $-2.00\pm0.00$ & 11   \\
1 & 18.0--20.0    & $(7.74\pm4.32)\times10^{-8}$ & $-4.55\pm1.81$ & 20   \\
2 & 20.0--23.0    & $(2.96\pm0.64)\times10^{-7}$ & $-5.44\pm0.89$ & 182  \\
3 & 23.0--25.0    & $(3.86\pm0.36)\times10^{-6}$ & $-2.72\pm0.17$ & 1942 \\ %\hline
4 & 25.0--30.0    & $(1.17\pm0.12)\times10^{-6}$ & $-3.76\pm0.24$ & 1424 \\
5 & 30.0--35.0    & $(5.35\pm0.80)\times10^{-7}$ & $-3.33\pm0.33$ & 548  \\ 
6 & 35.0--45.0    & $(3.56\pm0.49)\times10^{-7}$ & $-3.21\pm0.28$ & 621  \\
7 & 45.0--55.0    & $(2.35\pm0.40)\times10^{-7}$ & $-2.49\pm0.28$ & 387  \\
8 & 55.0--65.0    & $(9.16\pm2.13)\times10^{-8}$ & $-3.78\pm0.59$ & 137  \\
9 & 65.0--75.0    & $(7.45\pm1.72)\times10^{-8}$ & $-5.61\pm0.95$ & 155  \\
10& 75.0--85.0    & $(7.66\pm1.97)\times10^{-8}$ & $-3.35\pm0.57$ & 128  \\
11& 85.0--150.0   & $(3.48\pm0.53)\times10^{-8}$ & $-3.23\pm0.32$ & 280  \\
12& 150.0--300.0  & $(1.52\pm0.26)\times10^{-8}$ & $-2.99\pm0.32$ & 170  \\
13& 300.0--500.0  & $(8.63\pm1.72)\times10^{-9}$ & $-2.58\pm0.34$ & 94   \\
\hline
\end{tabular}
\end{table}

\section*{LLE comparison}
\begin{figure}
    \centering
    \includegraphics[width=\linewidth]{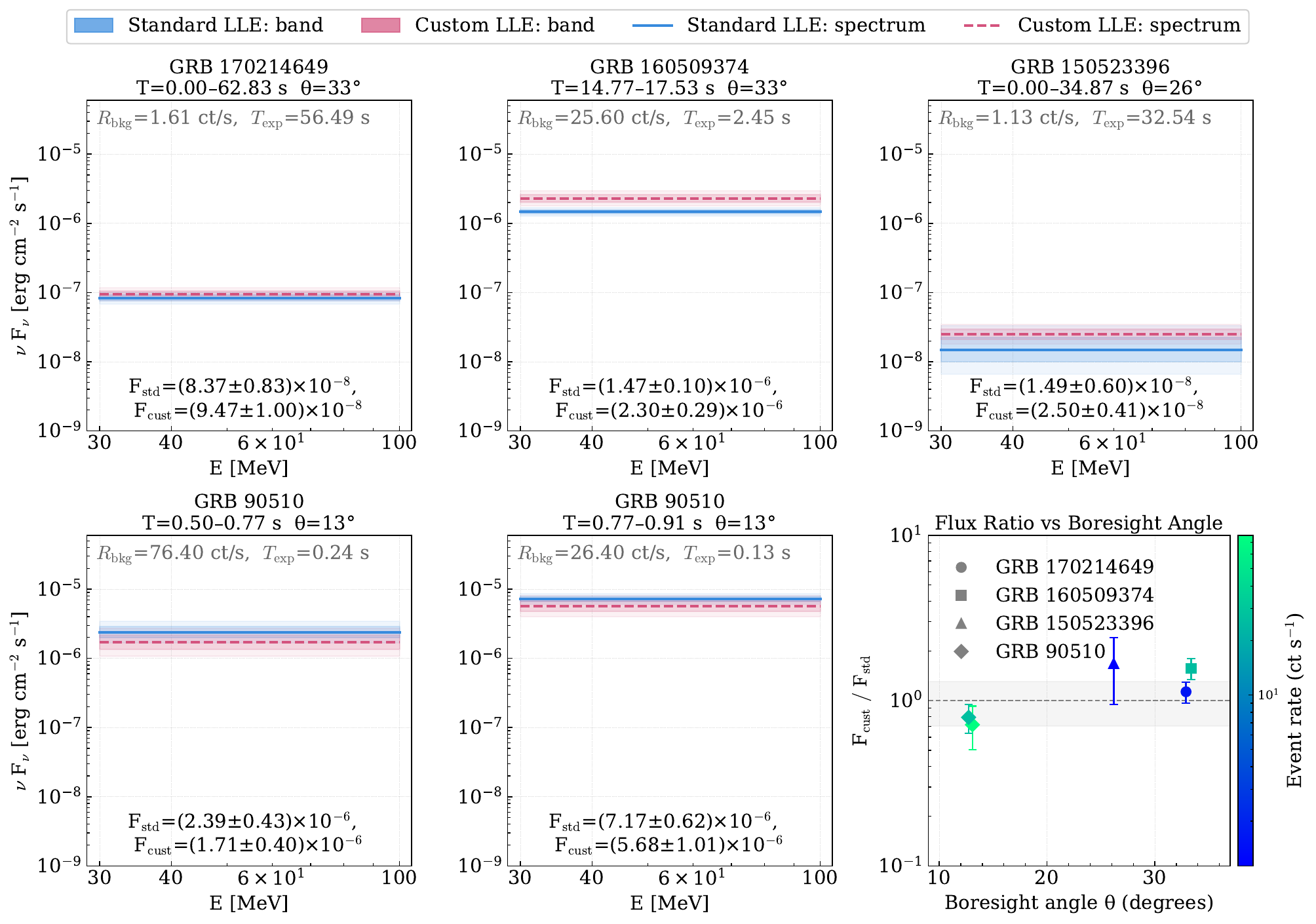}
\caption{Validation of the custom LLE spectral analysis pipeline against the standard LLE method for a sample of GRBs from \cite{Macera:2025wrv}, covering nearly two orders of magnitude in LLE flux ($\sim$30--100\,MeV). The first five panels show the $\nu F_{\nu}$ butterfly spectra for GRB~170214649 (medium flux, $\theta=33^{\circ}$), GRB~160509374 (high flux, $\theta=33^{\circ}$), GRB~150523396 (low flux, $\theta=26^{\circ}$), and two time bins of GRB~090510 (short GRB, $\theta=13^{\circ}$), with shaded bands indicating the $1\sigma$ (dark) and $2\sigma$ (light) confidence regions for the standard (blue) and custom (pink) pipelines. The spectral index is fixed to $\Gamma=-2$ in all fits. The sixth panel shows the flux ratio F$_{\rm cust}/{\rm F}_{\rm std}$ as a function of boresight angle $\theta$, with points colour-coded by background event rate (ct\,s$^{-1}$); different markers identify different GRBs. The grey band indicates the $\pm30\%$ agreement region, within which all measurements are consistent, demonstrating the reliability of the custom pipeline across the full dynamic range of the sample.}
    \label{fig:LLE_test}
\end{figure}
To validate the custom LLE spectral analysis pipeline described in Sect.~\ref{sec:LLE}, we selected a sample of GRBs seen in LLE from \cite{Macera:2025wrv}, spanning a broad range of LLE brightness in the energy range of $30-100$\,MeV. Specifically, we considered GRB~160509374 (high flux: $\sim10^{-6}$ erg\,cm$^{-2}$\,s$^{-1}$), GRB~170214649 (medium flux: $\sim10^{-7}$ erg\,cm$^{-2}$\,s$^{-1}$), and GRB~150523396 (low flux: $\sim10^{-8}$ erg\,cm$^{-2}$\,s$^{-1}$) as representative cases. In addition, we included GRB~090510, a short, hard GRB with two distinct time bins in the LLE band, providing a complementary test case with a very different spectral and temporal character. We note that GRB~090510 triggered LAT independently of MeV instruments. 

For each source, we performed a spectral analysis in the LLE energy range (30–100 MeV) using both the standard LLE pipeline and the custom method described above, extracting the energy flux with fixed index of $-2$. This assumption is valid since the energy band is narrow. The resulting spectral butterfly plots are shown in Fig.~\ref{fig:LLE_test}, where the 1$\sigma$ and 2$\sigma$ confidence bands are displayed for each method. 
To quantify the agreement, the sixth panel of Fig.~\ref{fig:LLE_test} shows the ratio F$_\mathrm{cust}/{\rm F}_\mathrm{std}$ as a function of boresight angle $\theta$, with each point colour-coded by the background event rate. The two methods agree within $\sim$30\% (the grey band in the right most plot in the lower panel of Fig.~\ref{fig:LLE_test}), consistent with systematic uncertainties in the LLE response. This cross-validation demonstrates that the custom LLE pipeline reproduces the standard analysis within the reported statistical uncertainties across nearly two orders of magnitude in LLE flux.

\begin{figure}
    \centering
    \includegraphics[width=1\linewidth]{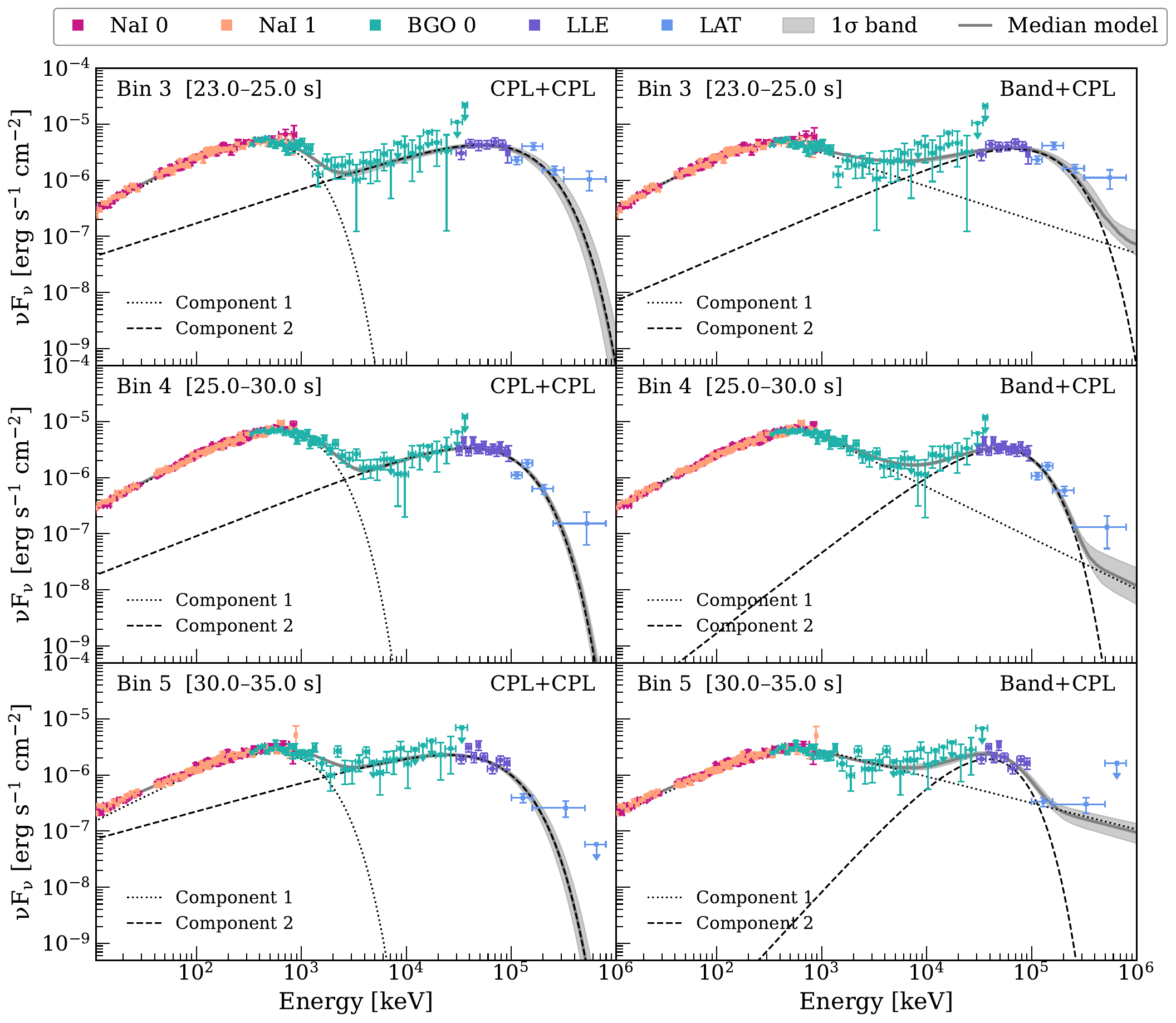}
    \caption{GRB 260226A spectral analysis during the transition phase, namely during bin 3 ($23-25$ s, first row), bin 4 ($25-30$ s, second row) and bin 5 ($30-32$ s, third row). We show the results of the CPL+CPL fits (left column) and Band+CPL fits (right column). In each spectrum, data points and $1\sigma$ error bars are shown as colored squares (NaI 0 in pink, NaI 1 in orange, BGO 0 in green, LLE in violet, LAT in light blue). $3\sigma$ upper limits are reported as arrows. The overall best fit model is represented with a solid line, while its $1\sigma$ error region with a gray band. We singularly show the contribution of both the first and second components with dotted and dashed lines, respectively.}
    \label{fig:Band/CPL+CPL_comparison}
\end{figure}

\section*{Spectral analysis tests}\label{sec:tests}

To assess the robustness of our results, we performed the time-resolved spectral analysis under several alternative configurations. In particular, we repeated the procedure described in Section~\ref{sec:spectral_fit_routine} by first removing the cross-calibration constant between the datasets, and then excluding both the calibration constants and the custom LLE dataset.

\begin{figure}
\centering
\includegraphics[width=0.8\textwidth]{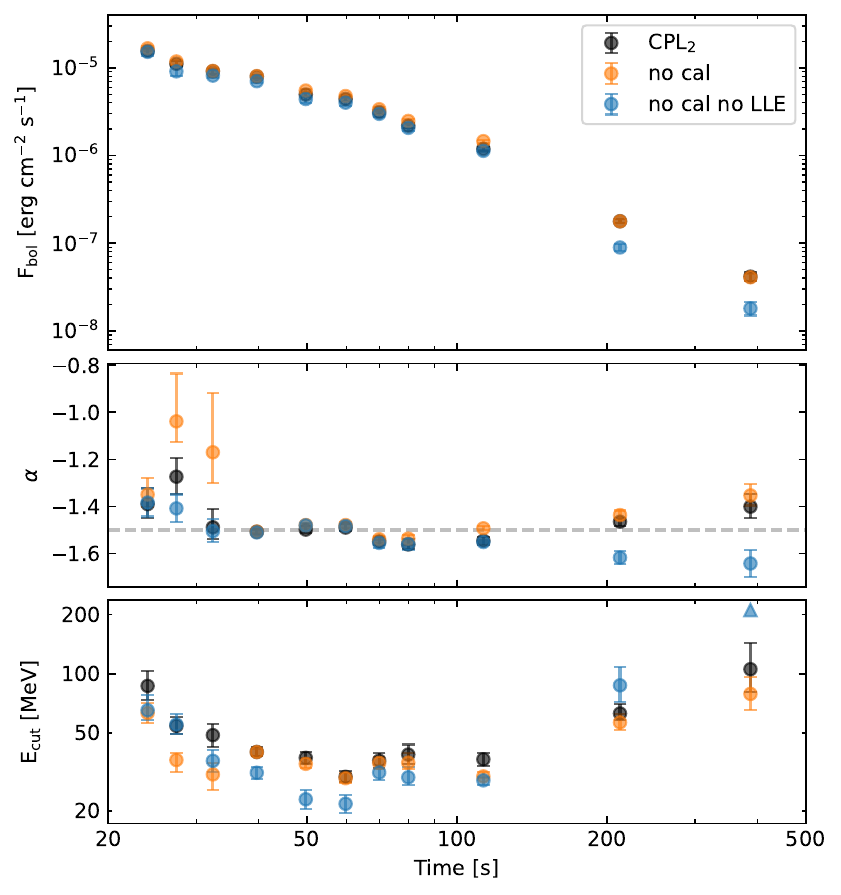}
\caption{Comparison of the temporal evolution of the second spectral component (bolometric flux, spectral index and cutoff energy of the CPL model), obtained using different analysis configurations. \\ Black points correspond to the results of the spectral analysis described in the main text. Orange points show spectral parameters obtained with the same setup but without applying calibration constants. Blue points indicate the results obtained when both the calibration constants are omitted and the LLE data are excluded from the spectral fits.}
\label{fig:spectral_analysis_comparison}
\end{figure}

The results of these tests are shown in Fig.~\ref{fig:spectral_analysis_comparison}, which presents the evolution of the spectral parameters of the second component. The black points correspond to our fiducial analysis, the orange points to the fit without cross-calibration constants, and the blue points to the case without both calibration constants and the LLE dataset. The results obtained with the three different configurations are consistent within the uncertainties, demonstrating the robustness of our findings.

Nevertheless, throughout this work we describe the second spectral component with a cutoff power law (CPL) at all epochs. This functional form is an empirical model rather than a physical model. Several features specific to the last two temporal bins (150--300 and 300--500 s) suggest that the apparent hardening of spectra might be related to systematic uncertainties. These are also the only two bins whose GBM spectra were extracted with the dedicated orbital-subtraction background model rather than the standard technique used elsewhere (Section \ref{sec:OSV_timebins}); the systematic uncertainty of this non-standard method has not been propagated into the fit, nor has the $\sim$40\% energy resolution of LLE\cite{Pelassa2010}. Finally, both bins have much lower photon statistics than the rest of the afterglow phase, and the last bin in particular carries a far broader relative uncertainty on $E_p$ than any earlier bin (Table \ref{fig:spectral_analysis_comparison}).

These caveats do not impact the observation of the ultrafast afterglow and, rather, strengthen the claim. Because $F_{\rm bol}$ and $E_p$ are correlated outputs of the same fit, an overestimated peak energy in these two bins implies an overestimated bolometric flux there as well. This effect is limited to the last two bins. If the correction tends to lower the peak energy, it will steepen the late-time decline relative to the $t^{-3}$ index reported in the main text.

Since it is difficult to precisely estimate the systematic uncertainties for the last two bins, we treat the peak energy as remaining within a conservative 15--60 MeV range (Fig. \ref{fig:second_component}) rather than fitting its temporal evolution, and note that this choice cannot be responsible for the extreme steepness of the reported decay.

\begin{table}
\caption{Time bins choice, fitted spectral models and model comparison results through the Akaike Information Criterion (AIC) relative to GRB 260226A time-resolved spectral analysis. For each tested model, we report the number of free parameters (not including the calibration constants), $k_0$. For the best-fit model of each time bin, we report the statistics, $stat$, and degrees of freedom, $dof$, of the fit.}
\centering
\begin{tabular}{c c c l c l c}
\hline
Bin & $t_{\rm start}$& $t_{\rm stop}$& Models[$k_0$] & AIC & Best-fit model &$stat/dof$\\
 & [s] & [s] &  &  &  \\
\hline

\multirow{2}{*}{0}  & \multirow{2}{*}{14}  & \multirow{2}{*}{18}  & Band[4]       & 485.08	 & \multirow{2}{*}{Band} & \multirow{2}{*}{$469.08/375$}\\
                    &                      &                      & BandCutoff[5]  & 486.42	 & \\
\hline

\multirow{2}{*}{1}  & \multirow{2}{*}{18}  & \multirow{2}{*}{20}  & Band[4]        & 562.36 & \multirow{2}{*}{BandCutoff} & \multirow{2}{*}{$446.05/374$}\\
                    &                      &                      & BandCutoff[5]  & 464.05	 & \\
\hline

\multirow{2}{*}{2}  & \multirow{2}{*}{20}  & \multirow{2}{*}{23}  & Band[4]        & 641.78 & \multirow{2}{*}{BandCutoff} & \multirow{2}{*}{$501.15/374$}\\
                    &                      &                      & BandCutoff[5]  & 519.15 & \\
\hline

\multirow{3}{*}{3}  & \multirow{3}{*}{23}  & \multirow{3}{*}{25}  & Band[4]                 & 540.62 & \multirow{3}{*}{CPL+CPL\footnotemark[1]}& \multirow{3}{*}{$477.73/373$} \\
                        &                      &                      & CPL+CPL[6]        & 497.73 & \\
                    &                      &                      & Band+CPL[7]    & 490.57 & \\
\hline

\multirow{3}{*}{4}  & \multirow{3}{*}{25}  & \multirow{3}{*}{30}  & Band[4]                 & 716.16 & \multirow{3}{*}{CPL+CPL\footnotemark[1]}& \multirow{3}{*}{$468.49/373$} \\
                    &                      &                      & CPL+CPL[6]        & 488.49 & \\
                    &                      &                      & Band+CPL[7]    & 478.85 & \\
\hline

\multirow{3}{*}{5}  & \multirow{3}{*}{30}  & \multirow{3}{*}{35}  & Band[4]                & 552.27 & \multirow{3}{*}{CPL+CPL\footnotemark[1]}& \multirow{3}{*}{$446.43/373$} \\
                    &                      &                      & CPL+CPL[6]        & 466.43 & \\
                    &                      &                      & Band+CPL[7]    & 458.85 & \\
\hline
6  & 35  & 45  & \rule{0pt}{3.2ex}CPL\footnotemark[2][3] & // & // & $436.92/376$ \\
\hline
7  & 45  & 55  & \rule{0pt}{3.2ex}CPL\footnotemark[2][3] & // & // & $443.20/376$ \\
\hline
8  & 55  & 65  & \rule{0pt}{3.2ex}CPL\footnotemark[2][3] & // & // & $438.04/376$ \\
\hline
9  & 65  & 75  & \rule{0pt}{3.2ex}CPL\footnotemark[2][3] & // & // & $435.85/376$ \\
\hline
10 & 75  & 85  & \rule{0pt}{3.2ex}CPL\footnotemark[2][3] & // & // & $353.48/376$ \\
\hline
11 & 85  & 150 & \rule{0pt}{3.2ex}CPL\footnotemark[2][3] & // & // & $341.22/376$ \\
\hline
12 & 150 & 300 & \rule{0pt}{3.2ex}CPL\footnotemark[2][3] & // & // & $699.11/486$ \\
\hline
13 & 300 & 500 & \rule{0pt}{3.2ex}CPL\footnotemark[2][3] & // & // & $705.18/486$ \\
\hline
\end{tabular}

\footnotetext[1]{Reported best-fit model varies from the one suggested by the Akaike Information Criterion (see Section~\ref{sec:comparison_2nd_comp})}
\footnotetext[2]{When a single model is tested, that is considered as the ``best-fit'' model.}
\label{tab:timebins_bestodels}
\end{table}

\begin{table}
\centering
\caption{Best-fit spectral parameters for each time bin of GRB 260226A. Fluxes $F$ in units of $10^{-6}$\,erg\,cm$^{-2}$\,s$^{-1}$. $E_{p}$ of Band and BandCutoff models in keV; $E_{cutoff}$ of BandCutoff model in MeV; $E_{p}$ of CPL model in MeV. CPL spectral indices follow the Band function convention ($\alpha = -\Gamma_{\rm CPL}$).}
\label{tab:bestfit_timeres}
\begin{tabular}{l c cccccc}
\toprule
& & \multicolumn{6}{c}{\textbf{Band}} \\
\cmidrule(lr){3-8}
Bin &  Interval  & $F$ & $\alpha$ & $E_{p}$ & $\beta$ & \multicolumn{2}{c}{} \\
& [s]& [$10^{-6}$\,cgs] & & [keV] & & & \\
\midrule
    0 & $(14,\,18)$ & $27.74_{-0.47}^{+0.48}$ & $-0.74_{-0.01}^{+0.01}$ & $679_{-14}^{+14}$ & $-3.12_{-0.06}^{+0.06}$ & & \\
\midrule
& & \multicolumn{6}{c}{\textbf{BandCutoff}} \\
\cmidrule(lr){3-8}
Bin & Interval & $F$ & $\alpha$ & $E_{p}$ & $\beta$ & $E_{cutoff}$ & \\
& [s]& [$10^{-6}$\,cgs] & & [keV] & & [MeV] & \\
\midrule
    1 & $(18,\,20)$ & $86.57_{-1.39}^{+1.40}$ & $-0.66_{-0.01}^{+0.01}$ & $601_{-12}^{+13}$ & $-2.45_{-0.03}^{+0.03}$ & $43_{-4}^{+6}$ & \\
    2 & $(20,\,23)$ & $66.46_{-1.12}^{+1.15}$ & $-0.76_{-0.01}^{+0.01}$ & $586_{-13}^{+13}$ & $-2.31_{-0.02}^{+0.02}$ & $59_{-7}^{+8}$ & \\
\midrule
& & \multicolumn{6}{c}{\textbf{CPL+CPL}} \\
\cmidrule(lr){3-8}
& & \multicolumn{3}{c}{CPL$_1$} & \multicolumn{3}{c}{CPL$_2$} \\
\cmidrule(lr){3-5} \cmidrule(lr){6-8}
Bin & Interval & $F$ & $\alpha$ & $E_{p}$ & $F$ & $\alpha$ & $E_{p}$ \\
& [s]& [$10^{-6}$\,cgs] & & [keV] & [$10^{-6}$\,cgs] & & [MeV] \\
\midrule
    3 & $(23,\,25)$ & $12.00_{-0.49}^{+0.47}$ & $-0.94_{-0.03}^{+0.03}$ & $417_{-16}^{+18}$ & $15.57_{-1.24}^{+1.25}$ & $-1.39_{-0.06}^{+0.07}$ & $53_{-6}^{+8}$ \\
    4 & $(25,\,30)$ & $17.24_{-0.44}^{+0.42}$ & $-0.94_{-0.01}^{+0.02}$ & $596_{-15}^{+16}$ & $11.12_{-0.72}^{+0.73}$ & $-1.27_{-0.07}^{+0.08}$ & $39_{-2}^{+3}$ \\
    5 & $(30,\,35)$ & $6.69_{-0.25}^{+0.37}$ & $-1.04_{-0.05}^{+0.05}$ & $529_{-14}^{+15}$ & $9.12_{-0.73}^{+0.64}$ & $-1.49_{-0.05}^{+0.08}$ & $25_{-2}^{+2}$ \\
\midrule
& & \multicolumn{6}{c}{\textbf{CPL}} \\
\cmidrule(lr){3-8}
Bin & Interval & $F$ & $\alpha$ & $E_{p}$ & \multicolumn{3}{c}{} \\
& [s]& [$10^{-6}$\,cgs] & & [MeV] & & & \\
\midrule
    6 & $(35,\,45)$ & $7.95_{-0.27}^{+0.28}$ & $-1.51_{-0.01}^{+0.01}$ & $20_{-1}^{+1}$ & & & \\
    7 & $(45,\,55)$ & $4.96_{-0.21}^{+0.24}$ & $-1.50_{-0.01}^{+0.01}$ & $19_{-1}^{+1}$ & & & \\
    8 & $(55,\,65)$ & $4.41_{-0.22}^{+0.22}$ & $-1.49_{-0.02}^{+0.01}$ & $15_{-1}^{+1}$ & & & \\
    9 & $(65,\,75)$ & $3.13_{-0.18}^{+0.19}$ & $-1.55_{-0.02}^{+0.02}$ & $16_{-1}^{+1}$ & & & \\
    10 & $(75,\,85)$ & $2.19_{-0.15}^{+0.16}$ & $-1.56_{-0.02}^{+0.02}$ & $17_{-1}^{+2}$ & & & \\
    11 & $(85,\,150)$ & $1.19_{-0.06}^{+0.06}$ & $-1.55_{-0.02}^{+0.02}$ & $17_{-1}^{+1}$ & & & \\
    12 & $(150,\,300)$ & $0.18_{-0.01}^{+0.01}$ & $-1.46_{-0.02}^{+0.02}$ & $34_{-2}^{+3}$ & & & \\
    13 & $(300,\,500)$ & $0.04_{-0.00}^{+0.00}$ & $-1.40_{-0.05}^{+0.06}$ & $63_{-11}^{+18}$ & & & \\
\bottomrule
\end{tabular}
\end{table}

\begin{table*}[htbp]
\centering
\caption{Time-resolved spectral analysis of GRB\,260226A (bins 3--5). Fluxes $F$ in units of $10^{-6}$\,erg\,cm$^{-2}$\,s$^{-1}$; $E_{\rm peak}$ of the soft component in keV, of the hard component in MeV. CPL spectral indices are in the \texttt{grbm} (Band) convention ($\alpha = -\Gamma_{\rm CutoffPL}$).}
\label{tab:spectral_timeres_2comp_comparison}
\begin{tabular}{l cccc ccc}
\toprule
& \multicolumn{7}{c}{\textbf{CPL+CPL}} \\
\cmidrule(lr){2-8}
& \multicolumn{4}{c}{Component 1 (CPL$_1$)} & \multicolumn{3}{c}{Component 2 (CPL$_2$)} \\
\cmidrule(lr){2-5} \cmidrule(lr){6-8}
Bin & $F$ & $\alpha$ & $E_{\rm peak}$ & & $F$ & $\alpha$ & $E_{\rm peak}$ \\
& [$10^{-6}$\,cgs] & & [keV] & & [$10^{-6}$\,cgs] & & [MeV] \\
\midrule
    3 & $12.00_{-0.49}^{+0.47}$ & $-0.94_{-0.03}^{+0.03}$ & $417_{-16}^{+18}$ & & $15.57_{-1.24}^{+1.25}$ & $-1.39_{-0.06}^{+0.07}$ & $53_{-6}^{+8}$ \\
    4 & $17.24_{-0.44}^{+0.42}$ & $-0.94_{-0.01}^{+0.02}$ & $596_{-15}^{+16}$ & & $11.12_{-0.72}^{+0.73}$ & $-1.27_{-0.07}^{+0.08}$ & $39_{-2}^{+3}$ \\
    5 & $6.69_{-0.25}^{+0.37}$ & $-1.04_{-0.05}^{+0.05}$ & $529_{-14}^{+15}$ & & $9.12_{-0.73}^{+0.64}$ & $-1.49_{-0.05}^{+0.08}$ & $25_{-2}^{+2}$ \\
\midrule
& \multicolumn{7}{c}{\textbf{Band+CPL}} \\
\cmidrule(lr){2-8}
& \multicolumn{4}{c}{Component 1 (Band)} & \multicolumn{3}{c}{Component 2 (CPL)} \\
\cmidrule(lr){2-5} \cmidrule(lr){6-8}
Bin & $F$ & $\alpha$ & $E_{\rm peak}$ & $\beta$ & $F$ & $\alpha$ & $E_{\rm peak}$ \\
& [$10^{-6}$\,cgs] & & [keV]& & [$10^{-6}$\,cgs] & & [MeV] \\
\midrule
    3 & $16.60_{-1.72}^{+1.88}$ & $-0.95_{-0.01}^{+0.01}$ & $400_{-12}^{+13}$ & $-2.60_{-0.19}^{+0.13}$ & $11.09_{-2.17}^{+2.08}$ & $-1.19_{-0.15}^{+0.26}$ & $69_{-9}^{+11}$ \\
    4 & $20.91_{-0.92}^{+0.86}$ & $-0.94_{-0.01}^{+0.01}$ & $591_{-16}^{+18}$ & $-2.91_{-0.17}^{+0.13}$ & $7.18_{-0.76}^{+1.05}$ & $-0.56_{-0.30}^{+0.29}$ & $50_{-3}^{+2}$ \\
    5 & $12.15_{-0.57}^{+0.50}$ & $-1.14_{-0.01}^{+0.01}$ & $592_{-22}^{+23}$ & $-2.47_{-0.07}^{+0.05}$ & $3.48_{-0.61}^{+0.82}$ & $0.05_{-0.50}^{+0.64}$ & $38_{-4}^{+4}$ \\
\bottomrule
\end{tabular}
\end{table*}

\begin{figure*}
    \centering
    \includegraphics[width=1\linewidth]{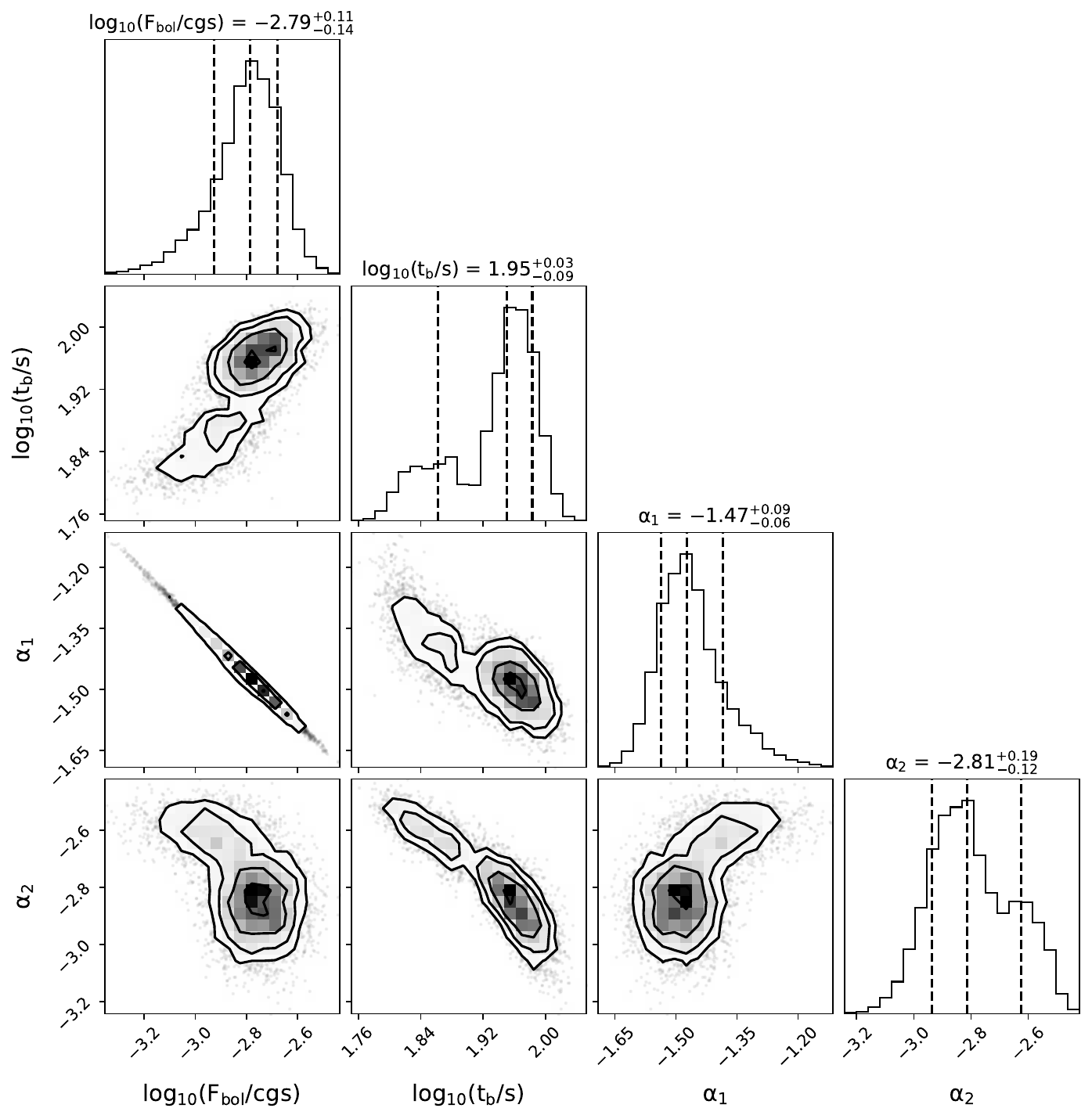}
    \caption{Posterior samples for the bolometric flux evolution of the second component, modelled with a broken power law. The dashed line in the middle represents the median of the distribution, whereas, the two dashed lines around represents the 16\% and 84\% containment value of the parameters. The contours represent the 1, 2 and 3 $\sigma$ containment-region for the 2 dimensional plots. }
    \label{fig:Fbol_posterior}
\end{figure*}

\begin{figure}[t]
  \centering
  \includegraphics[width=\linewidth]{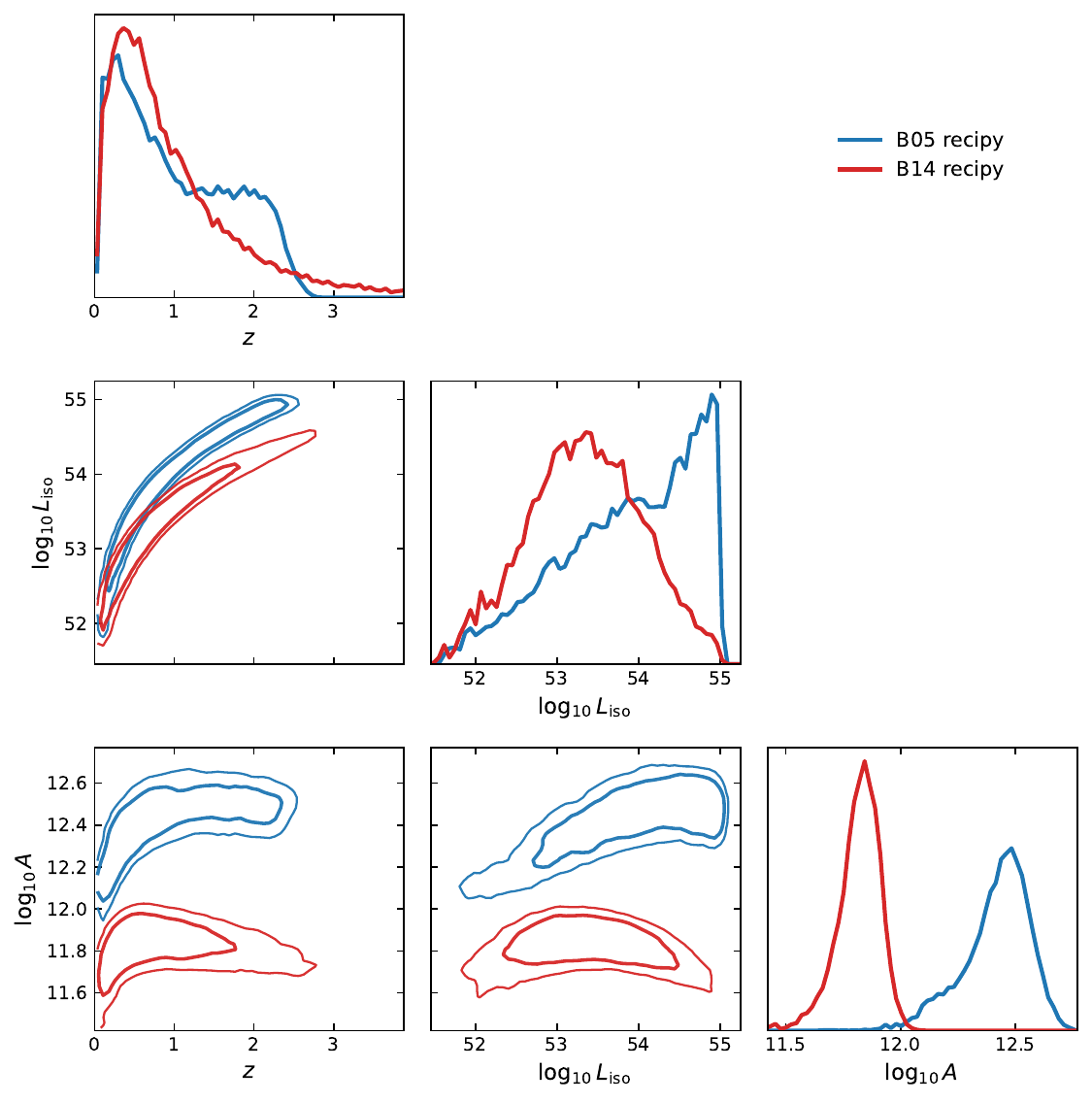}
  \caption{\textbf{Posterior distributions of the pair-loaded EIC model.}
  Joint posterior distributions of redshift ($z$), $\log_{10}L_{\rm GRB}$ and
  $\log_{10}A$ for the B05 cold-front and B14 hot-front pair-loading
  prescriptions.  Contours enclose 68\% and 90\% of the posterior
  probability.  The two prescriptions occupy different regions of parameter
  space, illustrating the systematic uncertainty associated with the radiation front
  pair-loading calculation.}
  \label{fig:eic_posteriors}
\end{figure}

\end{document}